\documentclass[12pt]{article}

\usepackage{amssymb}
\usepackage{epsfig} 
\usepackage{graphicx} 
\usepackage{color} 
\usepackage{times} 
\usepackage{amsmath} 
\usepackage{amssymb} 
\usepackage{xspace} 
\usepackage[varg]{txfonts} 
\usepackage{enumitem} 
\usepackage{natbib} 
\usepackage{algorithm} 
\usepackage{algorithmic} 
\usepackage{subfloat} 
\usepackage{subfig} 
\usepackage[normalem]{ulem}
\usepackage{booktabs}
\usepackage{lscape}
\usepackage{longtable}
\usepackage{hyperref}
\usepackage{authblk}
\usepackage[top=1in, bottom=1in, left=1in, right=1in]{geometry}

\usepackage{lineno}

\def\d3k{{\displaystyle {\rm d}{\bf k} \over \displaystyle (2\pi)^3}}

\def\apjl{Apj Letters} 
\def\aap{Astron. \& Astrophysics}

\def\apjl{Astrophysical Journal Letters}

\newcommand{\Excursion}  {\mm{{\mathbb E}}}
\newcommand{\Mask}       {\mm{{\mathsf M}}}
\newcommand{\Mspace}       {\mm{{\mathbb M}}}
\newcommand{\Rspace}     {\mm{{\mathbb R}}} 
\newcommand{\Sspace}     {\mm{{\mathbb S}}} 
 
\newcommand{\Homology}[1]{\mm{{\sf H}_{#1}}}
\newcommand{\Rank}[1]    {\mm{{\rm rank\,}{#1}}}
\newcommand{\Betti}[1]   {\mm{{\beta}_{#1}}}
\newcommand{\relBetti}[1]{\mm{{b}_{#1}}}
\newcommand{\Euler}      {\mm{\sf EC}} 
\newcommand{\relEuler}      {\mm{\sf EC_{\rm rel}}} 
 
\newcommand{\dime}[1]    {\mm{\rm dim\,}{#1}}
\newcommand{\ssx}        {\mm{\sigma}}
\newcommand{\tsx}        {\mm{\tau}}

\newcommand{\lips}{{\cal L}}

\newcommand{\R}{\mathbb{R}}

\newcommand{\x}{\mathbf{x}}

\newcommand{\y}{\mathbf{y}}

\newcommand {\mm}[1] {\ifmmode{#1}\else{\mbox{\(#1\)}}\fi}

\newcommand{\Res}            {\mm{N}}

\newcommand{\Skip}[1]        {}

\newcommand{\reltabNpipe}{
	
	\begin{tabular}{|r|r||r|r|r||r|r|r|} \hline
		
		\multicolumn{8}{|c|}{Relative homology} \\
		\hline
		
		&	& \multicolumn{3}{|c|}{Mahalanobis}  & \multicolumn{3}{|c|}{Tukey Depth} \\
		
		Res & \multicolumn{1}{|c||}{FWHM} & \multicolumn{1}{c}{$\relBetti{0}$} & \multicolumn{1}{c}{$\relBetti{1}$}
		
		& \multicolumn{1}{c}{$\relEuler$}
		
		& \multicolumn{1}{|c}{$\relBetti{0}$} & \multicolumn{1}{c}{$\relBetti{1}$}
		
		& \multicolumn{1}{c|}{$\relEuler$} \\ \hline \hline
		
		\multicolumn{8}{|c|}{threshold = 0.90} \\ \hline
		
		2048     &5   & 0.768 &	0.570&	0.858&		0.630&	0.318&	0.912 \\ 
		\hline
		
		1024   &10   & 0.780	& 0.599 &	0.383	&	0.672	& 0.352 &	\bf{0.000} \\
		\hline
		
		512  &20     & 0.280	& 0.549	& 0.203	&	0.332&	0.308&	\bf{0.000} \\
		\hline
		
		256   &40      & 0.585&	0.746&	0.772&		0.362&	0.560&	0.642\\
		\hline
		
		128    &80    & 0.120&	0.353&	0.563	&	\bf{0.000}&	0.312&	0.730 \\
		\hline
		
		64    &160    & 0.401	& 0.671 &	0.768	&	0.465&	0.718&	0.755 \\
		\hline
		
		32    & 320   & \bf{0.040}&	\bf{0.004}&	\bf{0.001}	&	\bf{0.000}&	\bf{0.000}&	\bf{0.000}\\
		\hline
		
		16    &640    & 0.987&	\bf{0.005}	&0.203	&	0.983&	\bf{0.000}&	0.440\\
		\hline
		
		summary&NA  & 0.478&	\bf{0.023}&	\bf{0.045}	&	0.803	& \bf{0.000}	& \bf{0.000}	  \\ 
		 \hline	\hline

	\end{tabular}
}

\newcommand{\reltabffp}{
	
	\begin{tabular}{|r|r||r|r|r||r|r|r|} \hline
		
		\multicolumn{8}{|c|}{Relative homology} \\
		\hline
		
		&	& \multicolumn{3}{|c|}{Mahalanobis}  & \multicolumn{3}{|c|}{Tukey Depth} \\
		
		Resolution & \multicolumn{1}{|c||}{FWHM} & \multicolumn{1}{c}{$\relBetti{0}$} & \multicolumn{1}{c}{$\relBetti{1}$}
		
		& \multicolumn{1}{c}{$\relEuler$}
		
		& \multicolumn{1}{|c}{$\relBetti{0}$} & \multicolumn{1}{c}{$\relBetti{1}$}
		
		& \multicolumn{1}{c|}{$\relEuler$} \\ \hline \hline
		
		\multicolumn{8}{|c|}{threshold = 0.90} \\ \hline
		
		2048     & 5 &0.454	& 0.573   	& 0.887		& \bf{0.000}	& 0.367	& 0.617\\ 
		\hline
		
		1024    & 10 & 0.728	 &    0.532	  &  0.690	&	0.580 &	\bf{0.000} &	\bf{0.000} \\
		\hline
		
		512      &     20 & 0.180 &	0.641 &	0.416	&	\bf{0.000} &	0.407 &	\bf{0.000} \\
		\hline
		
		256   &    40 &0.398 &	0.420 &	0.516	&	0.417 &	\bf{0.000} &	\bf{0.000}   \\
		\hline
		
		128    & 80 &0.075	&0.295	& 0.112	&	\bf{0.000} &	\bf{0.000} &	\bf{0.000} \\
		\hline
		
		64    & 160 &0.442 &	0.200 &	0.811	&	0.483 &	\bf{0.000}	& 0.720\\
		\hline
		
		32    & 320   & \bf{0.032}	& 0.192	& 0.140	&	\bf{0.000} &	0.353 &	\bf{0.000}		\\
		\hline
		
		16    &640    & 0.853 &	\bf{0.001} &	\bf{0.000}	&	0.870 &	\bf{0.000} &	\bf{0.000}	\\
		\hline
		
		summary&NA  & \bf{0.001}	& \bf{0.002}	 & \bf{0.000}	&	\bf{0.000} &	\bf{0.000} &	\bf{0.000}  \\ 
		\hline	\hline

	\end{tabular}
}

\begin{document}

\title{{Loops abound in the cosmic microwave background: A $4\sigma$ anomaly on super-horizon scales}}

\author{Pratyush Pranav}


\affil{Univ Lyon, ENS de Lyon, Univ Lyon1, CNRS, Centre de Recherche Astrophysique de Lyon UMR5574, F--69007, Lyon, France} 

\maketitle


\begin{abstract}
	We present a topological analysis of the temperature fluctuation maps from the \emph{Planck 2020} Data release 4 (DR4) based on the \texttt{NPIPE} data processing pipeline. For comparison, we also present the topological characteristics of the maps from \emph{Planck 2018} Data release 3 (DR3). We perform our analysis in terms of the homology characteristics of the maps, invoking relative homology to account for analysis in the presence of masks. We perform our analysis for a range of smoothing scales spanning sub- and super-horizon scales corresponding to $FWHM = 5', 10', 20', 40', 80', 160', 320', 640'$. Our main result indicates a significantly anomalous behavior of the loops in the observed maps compared to simulations that are modeled as isotopic and homogeneous Gaussian random fields. Specifically, we observe a $4\sigma$ deviation between the observation and simulations in the number of loops at $FWHM = 320'$ and $FWHM = 640'$, corresponding to super-horizon scales of $5$ degrees and larger. In addition, we also notice a mildly significant deviation at $2\sigma$ for all the topological descriptors for almost all the scales analyzed. Our results show a consistency across different data releases, and therefore, the anomalous behavior deserves a careful consideration regarding its origin and ramifications. Disregarding the unlikely source of the anomaly being instrumental systematics, the origin of the anomaly may be genuinely astrophysical -- perhaps due to a yet unresolved foreground, or truly primordial in nature. Given the nature of the topological descriptors, that potentially encodes information of all orders, non-Gaussianities, of either primordial or late-type nature, may be potential candidates. Alternate possibilities include the Universe admitting a non-trivial global topology, including effects induced by large-scale topological defects. 
\end{abstract}


\section{Introduction}
\label{sec:intro}

At the epoch of recombination, matter and radiation separate, allowing radiation to stream freely in the Universe. This free-streaming radiation permeating the Universe, that we observe as the \emph{Cosmic Microwave Background} (CMB) radiation, encodes a treasure trove of information about the initial conditions in the Universe \citep{ryden2003,jones2017precision}. Despite having a remarkably consistent average temperature, the CMB still exhibits tiny deviations of the order of $10^{-5}$ from the background average. The temperature fluctuations in the CMB trace the fluctuations in the underlying mater distribution in the infant Universe, that are linked to the spontaneous quantum fluctuations generated in an otherwise homogeneous medium  \citep{harrison1970,peebles1970}. Thus studying the properties of the temperature fluctuations in the CMB is essential towards understanding the properties of the primordial matter field. 

The Lambda Cold Dark Matter (LCDM) paradigm is the standard paradigm of cosmology. Together with the inflationary models in their simplest form \citep{starobinsky1982,guthpi1982}, the standard model of cosmology predicts the nature of the primordial stochastic matter distribution field to be that of an isotropic and homogeneous Gaussian random field \citep{harrison1970,guth1981}. This prediction finds allies theoretically in the Central Limit Theorem, and observationally in the various measurements of the CMB temperature anisotropy field via ground- and space-based probes such as the \texttt{BOOMERanG} balloon-based experiment \cite{boomerang} \citep{jaffe2001}, and the Wilkinson Microwave Anisotropy Probe (WMAP) satellite \citep{wmap9}. The latest endeavor of measuring the CMB temperature anisotropies has materialized through the launch of the \emph{Planck} satellite, which boasts of the highest resolution in measurements till date, at scales of a few arc-minutes \citep{planckOverview2018}. Despite the general consensus that the CMB exhibits the characteristics of an isotropic and homogeneous Gaussian random field, there is a growing body of evidence regarding anomalies in the observed CMB field with respect to the base model.  Notable among them are the observed hemispherical asymmetry in the CMB power spectrum \citep{eriksen2004}, as well as the alignment of low multipoles \citep{multipoles}; see \cite{cmbanomaliesstarkman} for a review. These observed anomalies raise doubts about the assumption of statistical isotropy and homogeneity respectively. 

Testing the assumption of Gaussianity requires tools which encode information about higher orders. Traditional endeavor in this direction has focused on  higher order correlation functions, which are generally extremely resource intensive computationally \citep{planckcollaboration2016a}. Recently, attention has turned towards developing alternative tools beyond the correlation functions and multi-spectra, which may potentially encode information of all orders. Principal such tools have arisen from integral geometry, and involve computing the \emph{Minkowski functionals} or the \emph{Lifshitz-Killing curvatures} \citep{adler1981,mecke94,schmalzinggorski,sahni1998,codis2013,ducout2013,matsubara2010,chingangbam2017,pranav2019a,eecestimate}. The $j$-th Minkowski functional and $(D-j)$-th Lifshitz-Killing curvature of a $D$-dimensional manifold $\Mspace$ are related by $Q_j(\Mspace) \ =\  j! \omega_j  \lips_{D-j}(\Mspace)$,   where $ j=0,\dots,D,$ and $\omega_j$ is the volume of the $j$-dimensional unit ball. There are $d$ such quantifiers for a $D$-dimensional set, where $d = 0, \ldots, D$. All but one are purely geometrical quantities, related to the $d$-dimensional volume of the manifold. The exception is the $0$-th Lifshitz-Killing curvature, or equivalently the $D$-th Minkowski functional, which is related to a purely topological quantity, the \emph{Euler characteristic} \citep{euler1758,adler1981}, via Gauss's \emph{Theorema Egrerium} \citep{gauss1900,adler1981,pranav2019a}. The Minkowski functional computations of the CMB have consistently shown the observations to be congruent with the standard model \citep{planckIsotropy2015}. 

More recently, developments in computational topology have paved way for extracting topological information from datasets, at the level of \emph{homology} \citep{munkres1984,edelsbrunnerharer10,isvd10,pranavthesis,pranav2017}, and its hierarchical extension \emph{persistent homology} \citep{edelsbrunnerharer10,pranav2017}. \emph{Topological data analysis} (TDA) involving homology and persistent homology has started finding application in astrophysical disciplines recently, for example in the context of structure identification \citep{shivashankar2015,xu2019} and quantification of large-scale structures \citep{pranavde,kono2020,wilding2020}, including detection and quantification of non-Gaussianities \citep{feldbrugge2019,biagetti2020}. Homology describes the topology of a space by identifying the holes and the topological cycles that bound them. A $d$-dimensional space may contain topological cycles of $0$ up to $d$-dimensions. The cycles and holes are associated with the \emph{homology groups} of the space. The $p$-th \emph{Betti number}, $\Betti{p}$, is the rank of the $p$-th homology group, $\mathbb{H}_{p, p = 0 \ldots d}$. While itself a purely topological quantity, the Euler characteristic is also the alternating sum of the Betti numbers of all ambient dimensions of a manifold, as denoted by the Euler-Poincar\'{e} formula \citep{adler1981,pranav2017,pranav2019a}. The Euler characteristic has a long history in the analysis of cosmological fields \citep{gdm86,pogosyan2009,ppc13,appleby2020}. 

Building on and refining existing tools from computational topology, in the context of analyzing the CMB field, this paper presents the homology characteristics of the temperature fluctuation maps of the cosmic microwave background obtained by the \emph{Planck} satellite \citep{planckOverview2018}. We perform our experiments on the fourth and the final data release \emph{Planck 2020} Data release 4 (DR4), which is based on the \texttt{NPIPE} data processing pipeline \citep{npipe}. The \texttt{NPIPE} dataset represents a natural evolution of the Planck data processing pipeline, integrating the best practices from the LFI and HFI pipelines separately. The result is an overall amplification of signal and reduction in the associated systematic, noise and residuals at almost all angular scales \citep{npipe}. For comparison, we also present results for the \emph{Planck 2018} Data release 3 (DR3) \citep{planckOverview2018}, which is based on the \emph{Full Focal Plane} (FFP) data processing pipeline \citep{plancksims} resulting in the \texttt{FFP10} simulations \citep{ffp10}. The paper follows the spirit of \cite{pranav2019b} in methods and analysis, where we present results for the intermediate \emph{Planck 2015} Data release 2 (DR2) \citep{planckIsotropy2015}. The novel aspect of the methodologies presented here and in \cite{pranav2019b} is an analysis pipeline that takes into account regions with unreliable data on $\Sspace^2$. In the case of CMB, this is reflected in the obfuscation effects that the measurements suffer from due to the foreground objects such as our own galaxy, as well as other extra- and intra-galactic foreground sources. We mask such regions and compute the homology of the excursion sets relative to the mask. 

Our main results indicate an anomalous behavior of loops when comparing the observational maps to the simulations. Specifically, we detect a $4\sigma$ deviation in the number of loops between the observational maps and simulations at scales of approximately $5$ degrees and larger. This is on top of a generally deviant behavior of both components and loops at around $2\sigma$ for almost all scales analyzed, when comparing observations to simulations. The Euler characteristic, being influenced by both the components and the loops, shows commensurate deviations. Even though differing in details, the results show a generally consistent trend across datasets. The results merit a serious consideration in view of the twin facts that the the data processing pipeline employed by the Planck team has evolved consistently, resulting in increasingly more accurate temperature maps, and that the temperature maps are in excellent agreement across data releases, not least due to a high signal-to-noise ratio a-priori. 

We present a brief description of the topological background in Section~\ref{sec:topology}, followed by the results in Section~\ref{sec:result}, which are based on the \texttt{NPIPE} dataset. We discuss the ramifications of the results and conclude the main body of the paper in Section~\ref{sec:discussion}. The appendices present a brief description of the datasets and the computational pipeline, as well as additional results based on the \texttt{FFP10} dataset.

\begin{figure}
	\centering
	\subfloat{\includegraphics[width=0.7\textwidth]{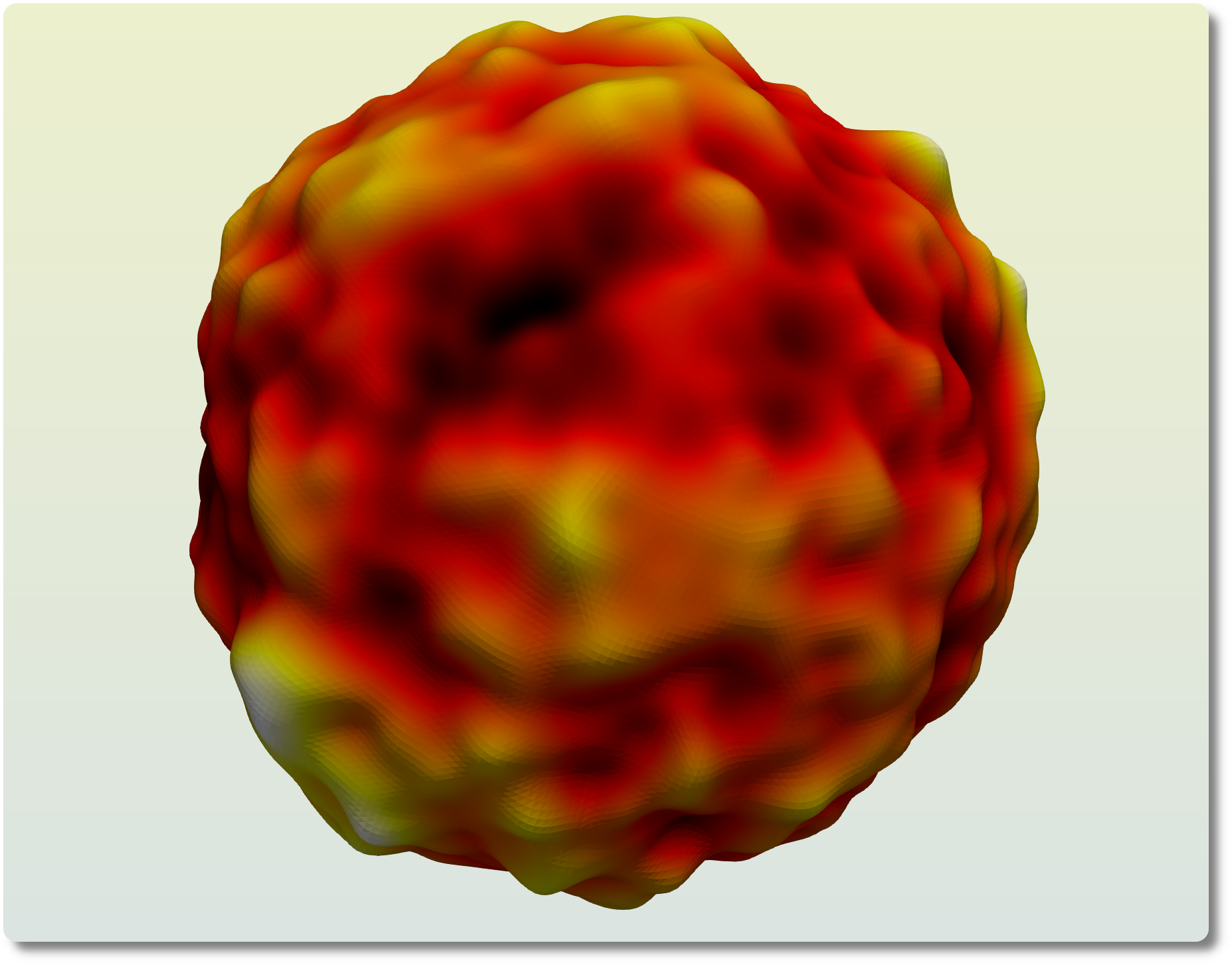} }
	\caption{A visualization of the temperature fluctuations in the CMB sky. The survey surface $\Sspace^2$ is distorted at each point in the direction of the surface normal. The distortion is proportional to the fluctuation in direction and magnitude. The visualization is based on the observed CMB sky cleaned by the \texttt{NPIPE} pipeline and smoothed at $5$ degrees.}
	\label{fig:cmbSky}
\end{figure}

\begin{figure*}
	\centering   
	\subfloat[]{\includegraphics[width=0.49\textwidth]{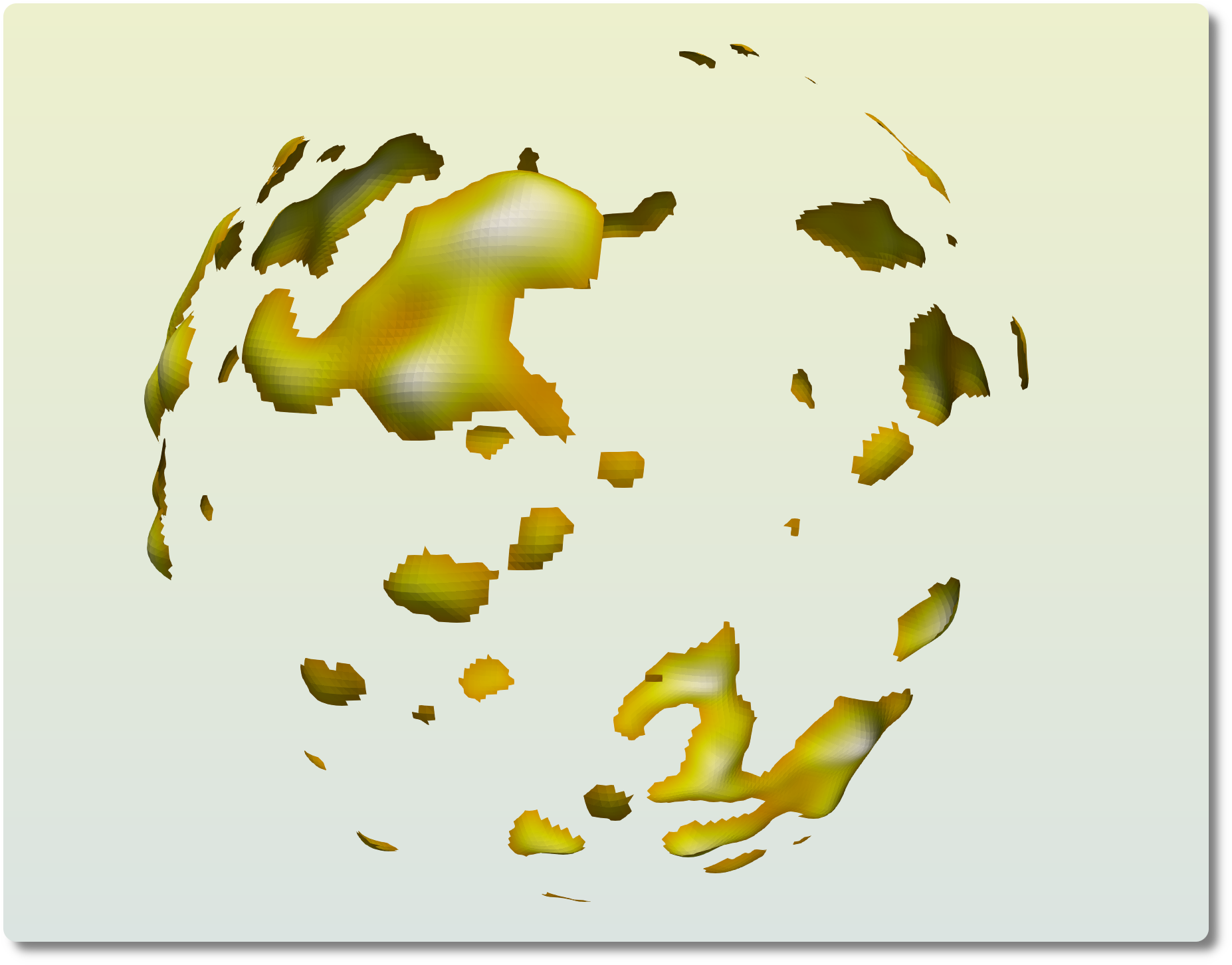}}
	\subfloat[]{\includegraphics[width=0.49\textwidth]{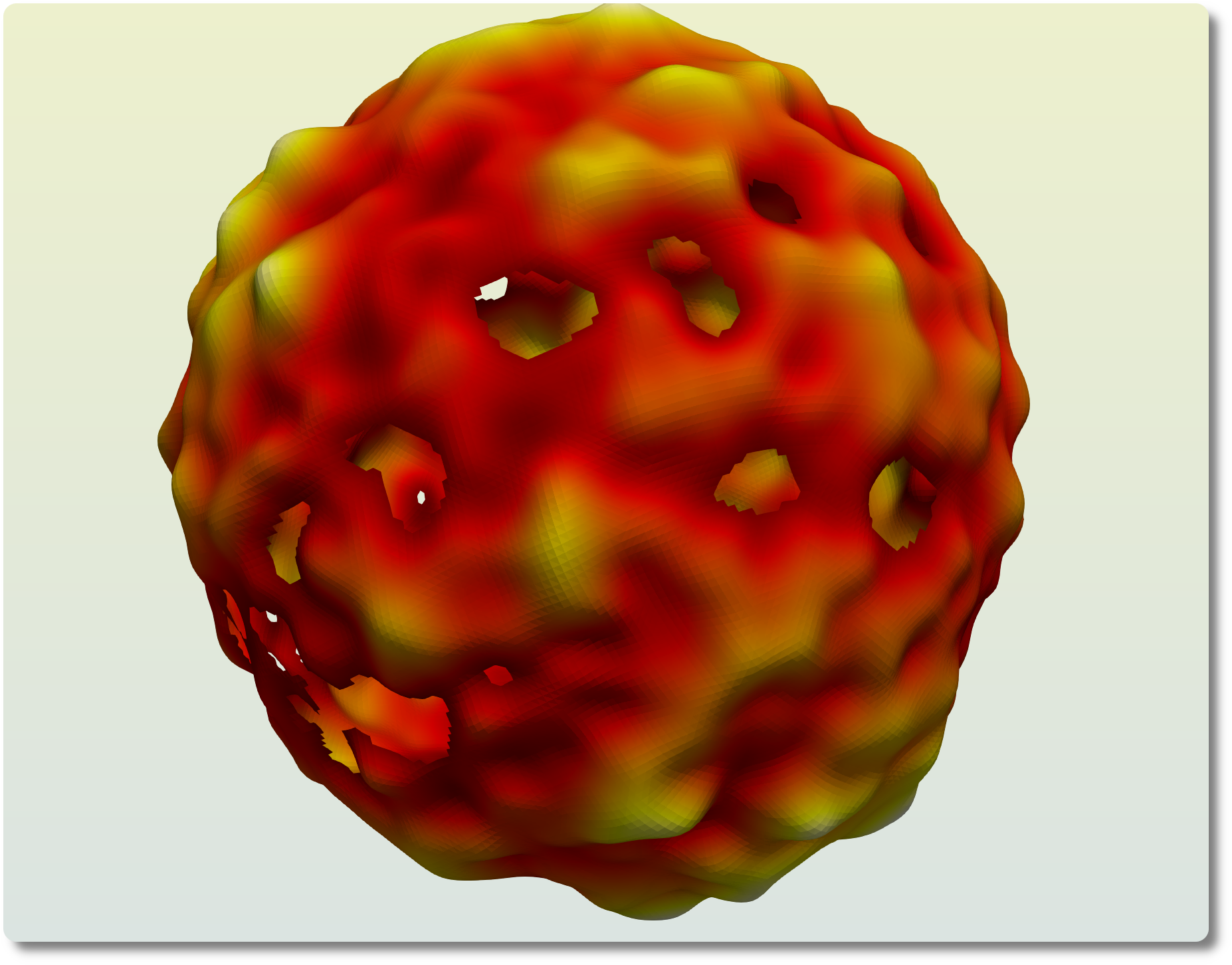}}
	
	\caption{The CMB sky thresholded at moderately positive (left) and negative (right) levels. For high thresholds, the excursion set is dominated by isolated components, while at low thresholds it gives the appearance of a single connected surface indented by numerous holes. For sufficiently low thresholds, the holes fill up, and the excursion set covers the entire sphere, composed of a connected surface without boundary, enclosing a single void (cf. Figure~\ref{fig:cmbSky}).}
	\label{fig:cmbThld}
\end{figure*}

\begin{figure}
	\centering
	\subfloat{\includegraphics[width=0.5\textwidth]{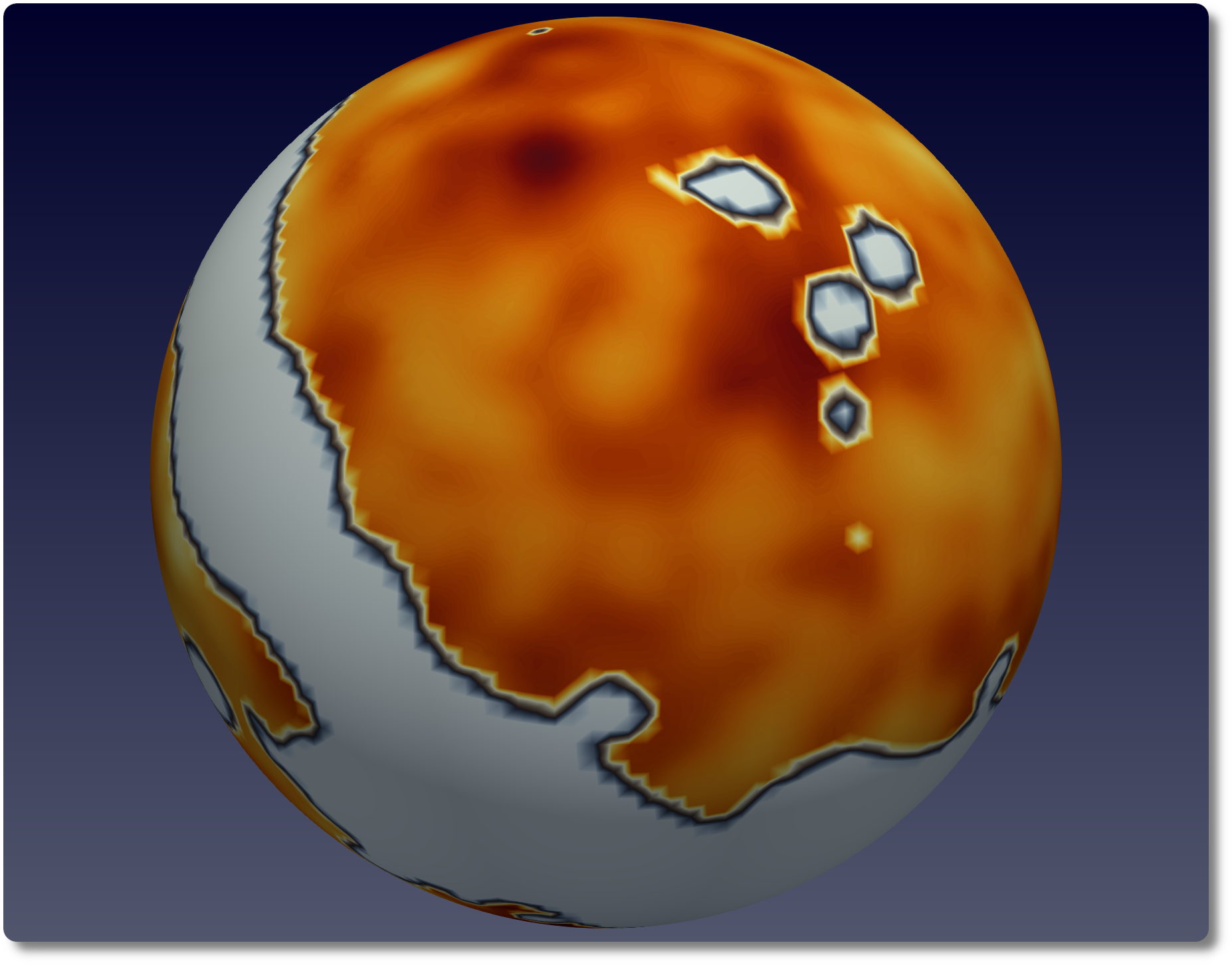} }
	\caption{A visualization of the temperature fluctuations in the CMB sky in the presence of masks (painted in grey). It consists of an equatorial belt and numerous patches on the northern and the southern cap, corresponding to our galaxy and other bright foreground objects. The visualization is based on the observed CMB sky cleaned by the \texttt{NPIPE} pipeline and smoothed at $5$ degrees.}
	\label{fig:maskedSky}
\end{figure}

\section{Topological background}
\label{sec:topology}

Commensurate with our intention of analyzing the topology of the CMB temperature fluctuations, we restrict ourselves to topological definitions on $\Sspace^2$ \citep{pranav2019b}, and invoke \emph{relative homology} to account for analysis in the presence of masked regions. Standard reference for this section is \cite{edelsbrunnerharer10}; also see \cite{pranav2019b} for discussion in the context of CMB.

\subsection{Homology characteristics of excursion sets of $\Sspace^2$}

Denoting the CMB temperature fluctuations on $\Sspace^2$ as $f \colon \Sspace^2 \to \Rspace$, we define the excursion set at a temperature $\nu$ as the subset of $\Sspace^2$ where the temperature is larger than or equal to $\nu$:
\begin{equation}
\Excursion (\nu)  =  \{ x \in \Sspace^2  \mid  f(x) \geq \nu \}.
\end{equation}

If $\Excursion (\nu)$ does not cover entire $\Sspace^2$, it may be composed of isolated components and holes. Figure~\ref{fig:cmbThld} presents excursion sets corresponding to two different thresholds. For high thresholds, presented in the left panel, the excursion set is dominated by components, while for low thresholds, presented in the right panel, the excursion set is dominated by a few large connected objects indented with holes, that are bounded by loops. The \emph{Betti numbers} $\Betti{0}$ and $\Betti{1}$ count the number of independent components and loops of the excursion set respectively. In general, for a $d$-dimensional topological space, $\Betti{p}$ is the rank of the $p$-th \emph{homology group}, $\Homology{p};p = 0, \ldots, d$, and counts the number of independent $p$-dimensional cycles \citep{munkres1984,edelsbrunnerharer10,pranav2017}. If $\Excursion (\nu)$ does not cover entire $\Sspace^2$, the number of independent loops is one less than the total number of loops. If $\Excursion (\nu)$ covers entire $\Sspace^2$, there are no loops, and $\Betti{2} = 1$, because of the void enclosed by the boundary-less surface of the sphere. A related quantity that has a long history of usage in cosmological analyses is the \emph{Euler characteristic}, or alternatively the \emph{genus} \citep{gdm86,ppc13}, which is the alternating sum of the Betti numbers of the excursion set:
\begin{equation}
\Euler (\nu) = \Betti{0} (\nu) - \Betti{1} (\nu) + \Betti{2} (\nu).
\end{equation}

The Euler characteristic also has a geometric interpretation as one of the \emph{Lifshitz-Killing} curvatures of the manifold \citep{adler1981,pranav2019b}.

\subsection{Masks and relative homology}

The measurement of CMB signal is unreliable in certain parts of the sky due to interference from bright foreground objects. These include extended objects such as our galaxy, as well as bright point sources. We mask such regions, and compute the homology characteristics of the excursion set relative to the mask. Figure~\ref{fig:maskedSky} presents a visualization of the masked CMB sky. Letting $\Mask \subseteq \Sspace^2$ be the mask, and $\Excursion (\nu)$ the excursion set,  we consider the \emph{relative homology} of the pair of closed spaces, $(E, M)$, where $E = \Excursion (\nu)$ and $M = \Mask \cap \Excursion (\nu)$. Note that $M$ is contained in $E$. We denote the rank of the \emph{relative homology groups} of the pair $(E, M)$ by $\relBetti{p} = \Rank{\Homology{p} (E, M)}; p = 0, 1, 2$. The Betti numbers computed considering the pair $(E, M)$ are different from the Betti numbers of the excursion set without a mask.  For a more detailed discussion about relative homology in the context of masked CMB sky see \cite{pranav2019b}. The \emph{relative Euler characteristic}, as in the case of absolute homology, is the alternating sum of the rank of relative homology groups:
\begin{equation}
\relEuler (\nu) = \relBetti{0} (\nu) - \relBetti{1} (\nu) + \relBetti{2} (\nu).
\end{equation}

\section{Results}
\label{sec:result}

We present our results in terms of the ranks of relative homology groups, $\relBetti{p}$ for $0 \leq p \leq 1$. 
We present the graphs of $\relBetti{0}$, $\relBetti{1}$, and of the (relative) Euler characteristic, $\relEuler$, followed by statistical tests that estimate the significance of results. The main paper presents results from the \texttt{DR4 NPIPE} dataset, based on $600$ simulations, obtained using the \texttt{SEVEM} component separation pipeline. Similar results for the  the \texttt{DR3 FFP10} dataset, based on $300$ simulations,  obtained using the \texttt{SMICA} component separation pipeline, are presented in the appendix for comparison.

\begin{figure}
	\centering
	\subfloat{\includegraphics[width=0.7\textwidth]{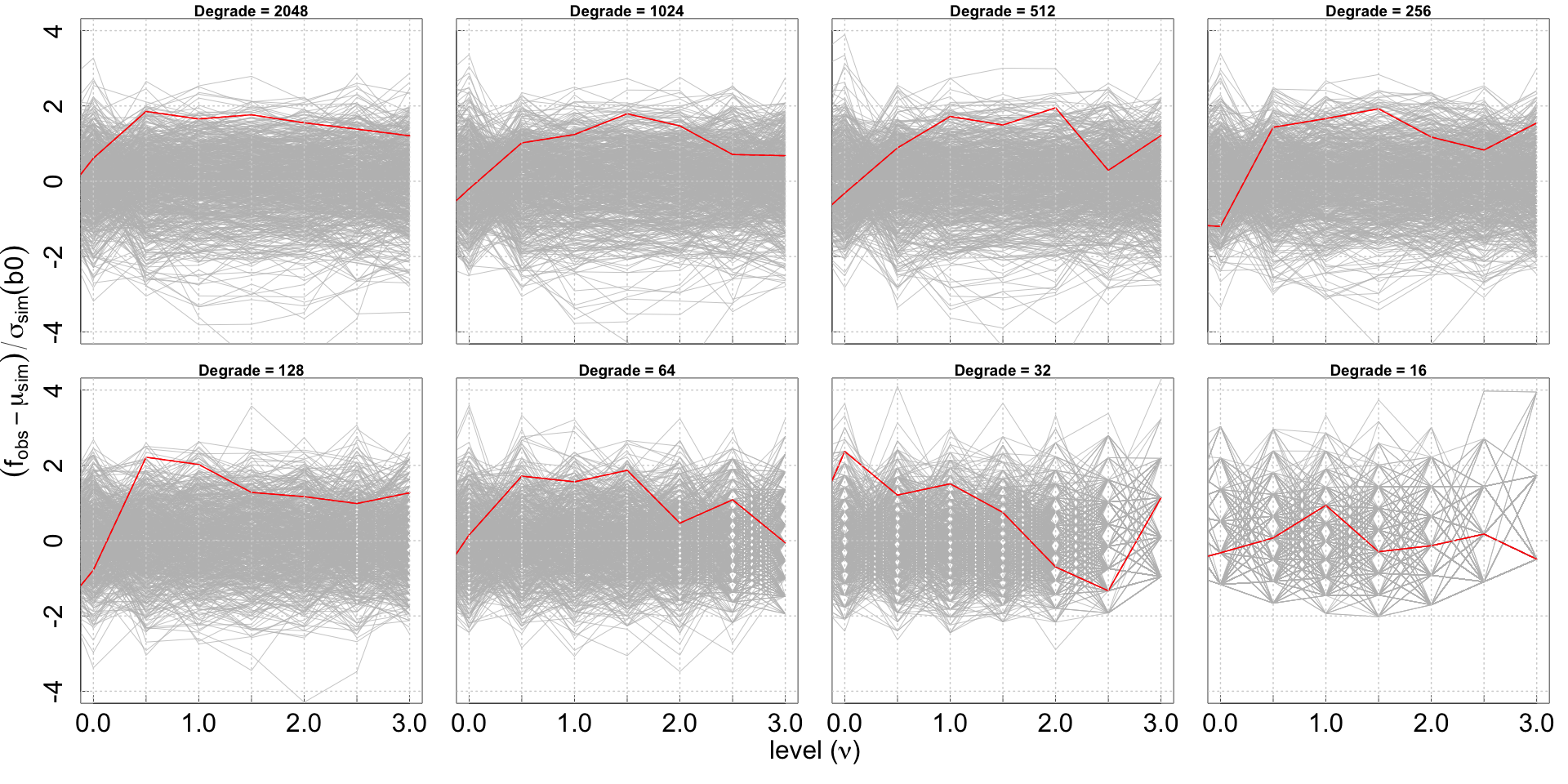} }\\
	\subfloat{\includegraphics[width=0.7\textwidth]{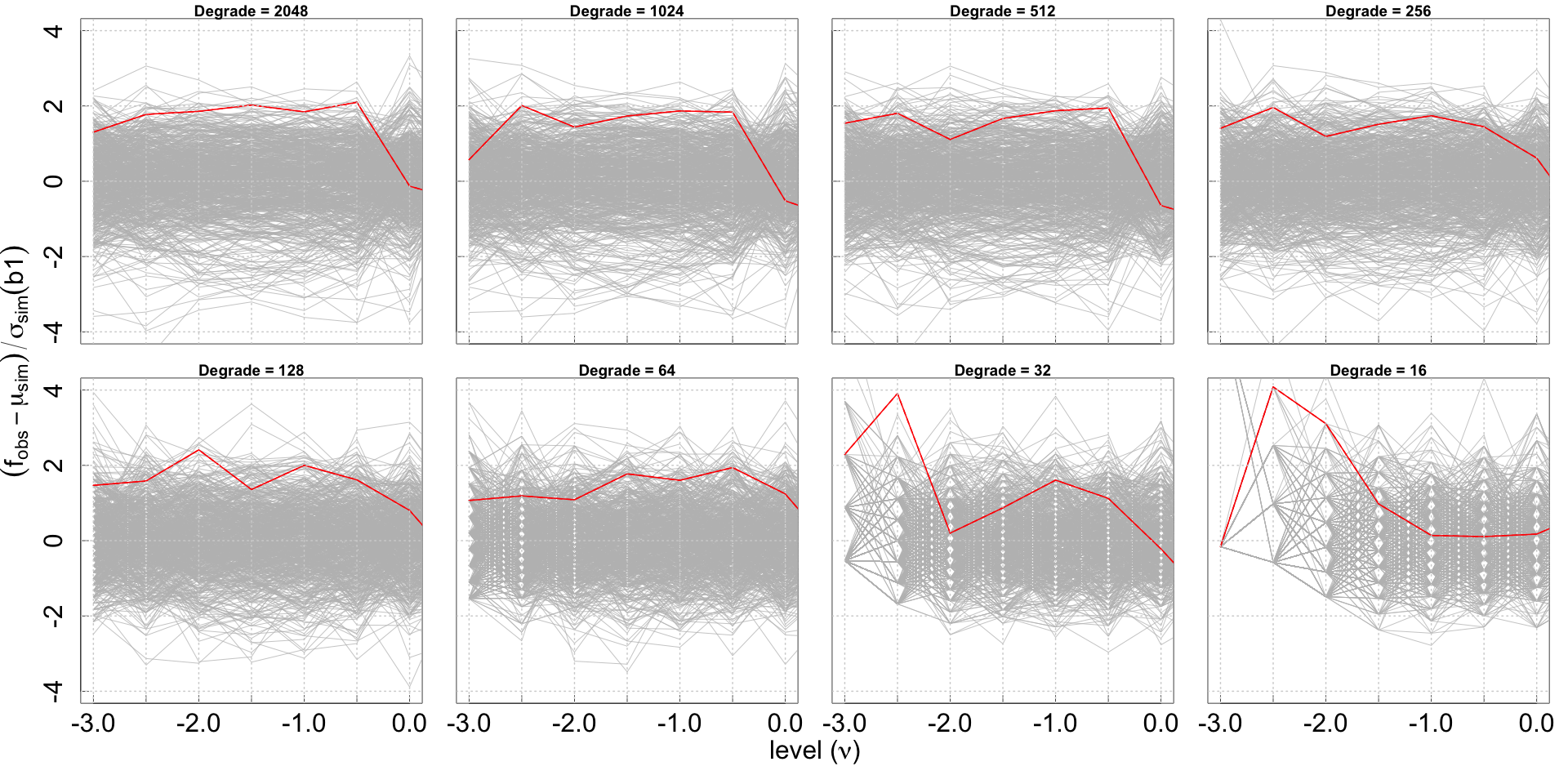} }\\
	\subfloat{\includegraphics[width=0.7\textwidth]{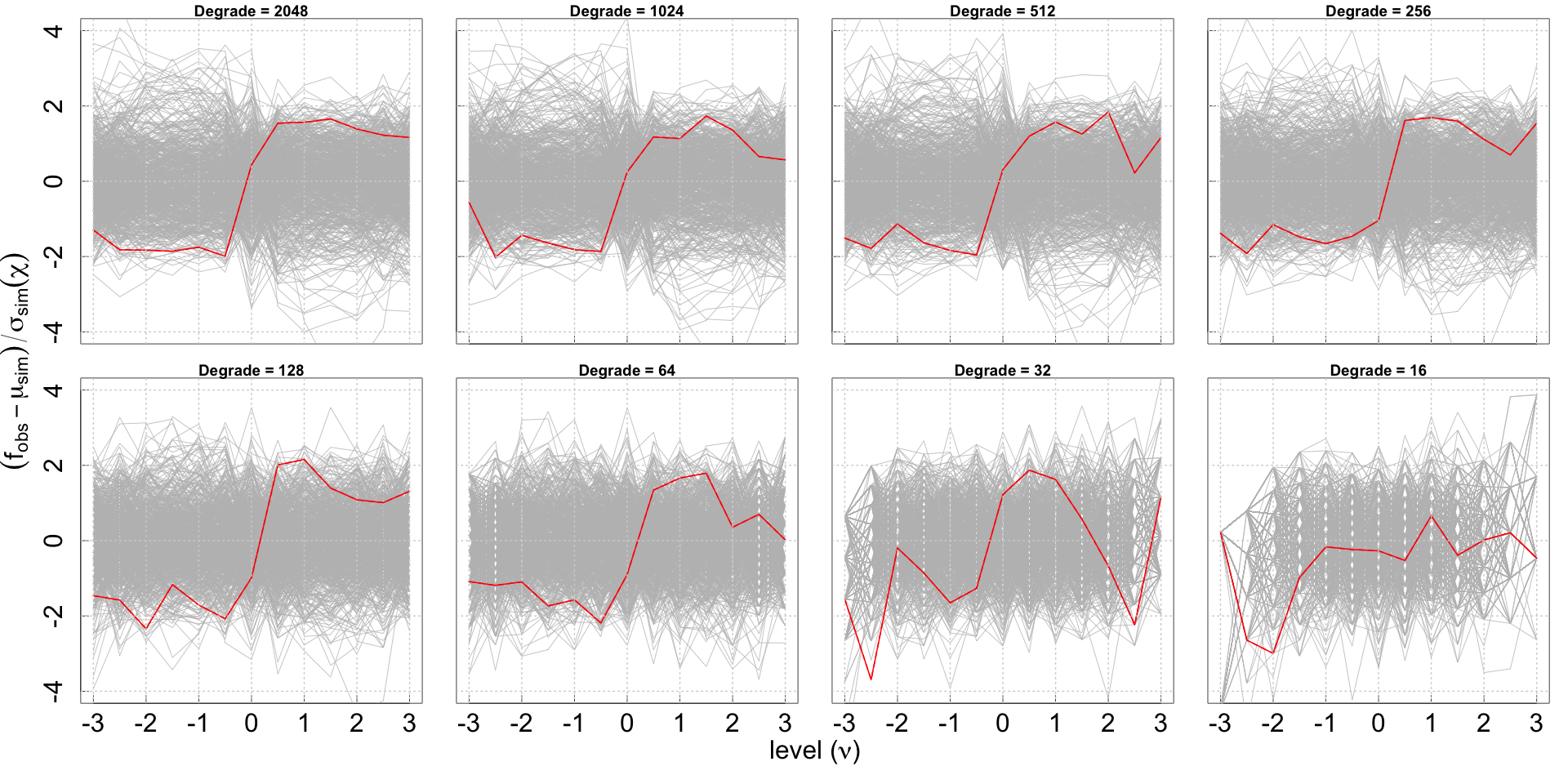} }\\
	\caption{Significance of difference for $\relBetti{0}$, $\relBetti{1}$ and $\relEuler$. The observational curve is presented in red, and the curves corresponding to simulations are presented in gray. The plotted quantity is the difference between the observation and the mean, scaled by the  standard deviation, the latter two computed from the simulations. We notice a $4\sigma$ deviation between observation and simulations in the number of loops at $\Res = 32$, and $\Res = 16$, corresponding to $FWHM = 320'$ and $FWHM = 640'$. This is on top of a $2\sigma$ deviation that we notice generally for both components and loops across almost all resolutions. The Euler characteristic reflects the deviations in components and loops, albeit at smaller significance, due to cancellation effects \citep{pranav2019b}.}
	\label{fig:nrm_graph}
\end{figure}

\begin{table}
	\tabcolsep=0.09cm
	\centering
	\subfloat{\reltabNpipe}
	\caption{Table displaying the two-tailed $p$-values for relative homology obtained from parametric (Mahalanobis distance) and non-parametric (Tukey depth) tests, for different resolutions and smoothing scales for the \texttt{NPIPE} dataset. The last entry is the $p$-value for the summary statistic computed across all resolutions. Marked in boldface are $p$-values $0.05$ or smaller.} 
	\label{tab:npipe_degrade-pvalues}
\end{table}

\begin{figure}
	\centering
	\subfloat{\includegraphics[width=0.7\textwidth]{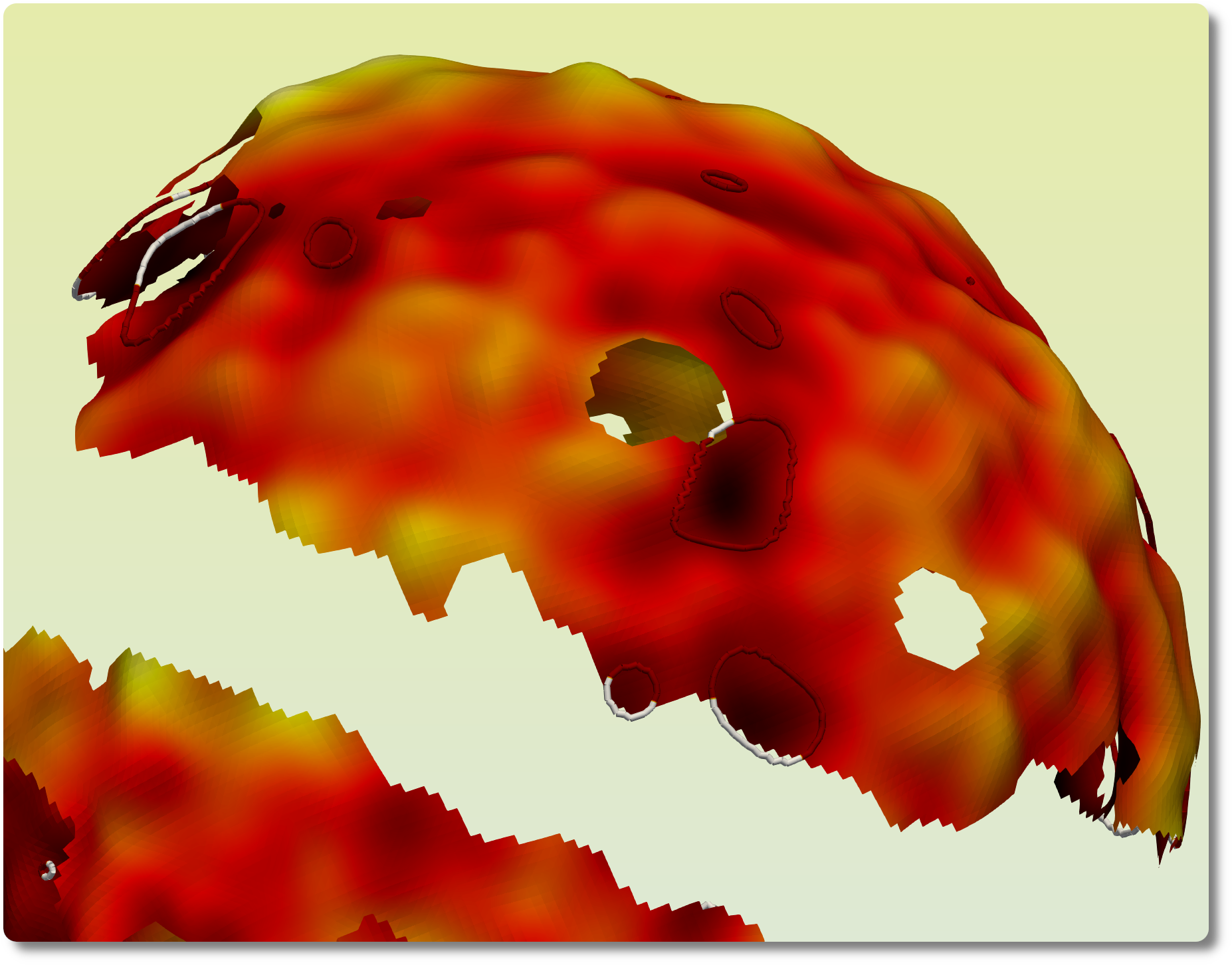} }\\
	\caption{A visualization of the loops surrounding the low density regions at a moderately negative threshold. The gaps in the manifold correspond to the mask that has been removed. We clearly notice the equatorial belt corresponding to our galaxy, and more patches in the visible cap. Some loops live fully in the excursion region, not influenced by the mask. We color them in red. We also depict some representative relative loops, that have their end points in mask. We draw closed loops for this category as well, coloring the portions in white where they overlap with the masked regions. The visualization is based on the observed CMB map from the \texttt{NPIPE} data set smoothed at $5$ degrees -- approximately the scale at which the loops start showing deviant behavior.}
	\label{fig:loops}
\end{figure}

\subsection{Graphs of $\relBetti{0}$, $\relBetti{1}$ and $\relEuler$}

 We choose a number of  a-priori levels, $\ell_{k}$, where 
$\ell_k = k/2$, setting $k_{\relBetti{0}} = [0:6]$,  $k_{\relBetti{1}} = [-6:0]$, and $k_{\relEuler} = [-6:6]$. We do so to restrict ourselves to analyzing $\relBetti{0}$ for positive thresholds, $\relBetti{1}$ for negative thresholds, and $\relEuler$ across the full threshold range. The choice of regions is determined by the fact that $\relBetti{0}(\nu)$  tends to be  small, and carries little information for $\nu<0$, $\relBetti{1}(\nu)$ tends to be small for $\nu>0$, while the Euler characteristic is informative over the full range of levels. We consider collections of random variables 
$\relBetti{0}(\ell_k)$, $\relBetti{1}(\ell_{k})$, and $\relEuler(\ell_{k})$. For each level, we compute the mean $\mu_{sim}$ and  the standard deviation $\sigma_{sim}$ from the simulations, for each of the topological descriptor. If the observed value is $f_{obs}$, the normalized difference, $\Delta$ is given by:
\begin{equation}
\Delta = \frac{f_{obs}-\mu_{sim}}{\sigma_{sim}}.
\end{equation}

\noindent Figure~\ref{fig:nrm_graph} presents the curves for the normalized differences for $\relBetti{0}$ (top two rows), $\relBetti{1}$ (middle two rows), and $\relEuler$ (bottom two rows). For each quantity, from top-left to bottom-right, we present plots corresponding to the $8$ different resolutions and smoothing scales enumerated in Table~\ref{tab:unmasked_area}. The normalized difference curves for the observational map is presented in red, overlapped with similar curves for the simulations,  presented in gray. The first thing we notice is that the observational curves for both $\relBetti{0}$ and $\relBetti{1}$ show mildly significant deviation at approximately $2\sigma$ across resolutions, for at least some thresholds. A more interesting observation is the highly significant deviation of the observational curves corresponding to relative loops at $4\sigma$ for $\relBetti{1}$, at $\Res = 32$ and $\Res = 16$, corresponding to $FWHM = 320'$ and $FWHM = 640'$. $\relBetti{1}$ quantifies the number of independent loops in the manifold. Figure~\ref{fig:loops} presents a visualization of some such loops at moderately negative thresholds, for observational maps smoothed at $5$ degrees. The smoothing scale is approximately the scale at which the loops start exhibiting a deviant behavior from the base model. The relative Euler characteristic also shows deviations for these resolutions, albeit at a slightly smaller significance, owing to cancellation effects \citep{pranav2019b}.  However, the Euler characteristic is not strictly an independent quantity, due to it being an alternating sum of the ranks of the relative homology groups. As such, it merely reflects the deviations in the contributing Betti numbers. For comparison, we also show graphs of the topological quantities computed from the \texttt{DR3 FFP10} dataset in Figure~\ref{fig:nrm_graph_ffp}. We notice similar trends in the behavior of the three topological descriptors as the case presented in the main body. 

\subsection{Statistical significance of the results}

We consider the two methods detailed in Section~\ref{sec:stat}, and
present $p$-values of the observed maps for both. Considering all the three topological quantities $\relBetti{0}$, $\relBetti{1}$ and $\relEuler$ we first compute the statistics combining all resolutions, in order to probe for non-random deviations. Thereafter, we compute the statistics for all resolutions separately, to ascribe a scale-dependence to the signals.

\medskip\noindent\textbf{Summary tests.} To test for the evidence of a non-random discrepancy, we take the full set of measurements for the relevant thresholds, combined across resolutions, for the three variables. This results in a data vector of $m = 56$ entries for   $\relBetti{0}$ and $\relBetti{1}$, and $m = 104$ for $\relEuler$. The length of the vector is commensurate with the fact that we analyze $\relBetti{0}$ and $\relBetti{1}$ for positive and negative thresholds respectively, while we analyze $\relEuler$ for the full range of levels. The last entry in the Table~\ref{tab:npipe_degrade-pvalues} presents the $p$-values for the variables for both the parametric and the non-parametric tests. Overall, there is an indication that the observations are deviant from the simulations, at least in the the number of loops, which is an independent quantity. By association, Euler characteristic also shows deviations. We also note that the non-parametric test shows higher significance compared to the parametric test. Table~\ref{tab:ffp10_degrade-pvalues} in the appendix presents the $p$-values for the \texttt{FFP10} dataset. The trends are similar, albeit with higher significance. The Mahalanobis test indicates stronger deviation for the \texttt{FFP10} dataset compared to the \texttt{NPIPE} dataset. Similarly, the Tukey depth test exhibits more instances of highly significant deviations for all the three variables, where the number of simulations are not enough to resolve the $p$-values. 

\medskip\noindent\textbf{Tests at specific resolutions.} All entries before the last in Table~\ref{tab:npipe_degrade-pvalues} present $p$-values for the variables, computed from maps degraded at specific resolutions detailed in Table~\ref{tab:unmasked_area}. Considering the Mahalanobis distance, $\relBetti{1}$ shows highly significant difference between the observational maps and the simulations for $\Res = 32$ and $\Res = 16$. Additionally, $\relBetti{0}$ also shows mildly significant difference at $\Res = 32$, but the differences are an order smaller than what is exhibited by $\relBetti{1}$. The Euler characteristic shows significant difference at $\Res = 32$. This is due to the significant difference shown by the contributing $\relBetti{0}$ and $\relBetti{1}$ at this resolution. In general, due to the Euler characteristic being an alternating sum of the Betti numbers, its behavior is influenced by both the Betti numbers. As an example, for the next lower resolution, $\Res = 16$, the Euler characteristic shows no significant difference, even though $\relBetti{1}$ exhibits significant difference. This is because of the highly non-significant behavior of $\relBetti{0}$, whose contribution ends up canceling the contribution of $\relBetti{1}$  towards the Euler characteristic. Results for the \texttt{FFP10} dataset are presented in Table~\ref{tab:ffp10_degrade-pvalues} in the appendix. As in the summary test case, the  Mahalanobis test indicates stronger deviation for the \texttt{FFP10} dataset compared to the \texttt{NPIPE} dataset in general. The Tukey depth test exhibits more instances of highly significant deviations across different resolutions for all the three variables, where the number of simulations are not enough to resolve the $p$-values.

\section{Discussions and conclusion}
\label{sec:discussion}

We present a topological analysis of the temperature fluctuations  in the CMB in terms of homology. To account for regions with unreliable data, we compute the homology of the excursion sets relative to the mask covering these areas. We perform our analysis on the fourth and final \texttt{NPIPE} data release from the Planck team. The pipeline represents a natural evolution of the data processing pipeline, commensurate with better understanding of systematics, residuals and noise over a period of time across successive data releases. It incorporates the best strategies from both the LFI and HFI processing pipeline, so that there is an overall reduced level of noise and residuals across all scales \citep{npipe}. We also investigate the \emph{Planck 2018} Data release 3 (DR3), accompanied by the  \texttt{FFP10} simulations \citep{ffp10}, for comparison and completeness. The present paper is a successor to \cite{pranav2019b}, where we investigate the topology of the temperature fluctuation maps  based on the intermediate  \emph{Planck 2015 } Data release 2 (DR2), accompanied by \texttt{FFP8} simulations. Between the two papers, we have investigated topological characteristics of the CMB temperature fluctuation maps for the latest three data releases by the Planck team, and the overall trends in the results across the datasets inform on the consistency of the data processing pipeline, as well as our own methodologies. 

Our main result indicates an anomaly in the behavior of the loops in the observed sky compared to the simulations based on the standard LCDM model that predicts the nature of the primordial perturbations to be that of an isotropic and homogeneous Gaussian random field. The number of loops in the observational maps shows significantly anomalous deviations from the simulations at $4\sigma$, at scales of $5$ degrees and larger. We also notice a general mild flaring of all the topological quantities computed from the observational maps at approximately $2\sigma$ for almost all resolutions. In order to test for non-random discrepancies, we compute the $p$-values using the parametric $\chi^2$ test, as well as the non-parametric Tukey depth test. For the parametric test, the number of loops shows per-mil deviations between observations and simulation, which is an order of magnitude larger than the deviation shown by the number of components. Non-parametric tests exhibit trends that are starker, more often for the loops compared to the other quantities,  in the sense that the number of simulations turn out to be inadequate to resolve the $p$-values reliably. The trend in our results show a consistency across the latest data releases \texttt{NPIPE} and \texttt{PR3} analyzed in this paper, as well as the intermediate data release \texttt{PR2} analyzed in \citep{pranav2019b}. The trend is commensurate with the general observation by the Planck team that the quality of the temperature maps has been consistent since the second data release PR2 \citep{planckIsotropy2015}. The consistency of our results across the three datasets independently confirms this. 

The origin and repercussions of the observed anomaly deserve careful consideration. Concerning the origin of the anomaly, the  least interesting, and a highly unlikely possibility is instrumental systematics as source. This is considering the twin facts that the results are consistent across data releases and different processing pipelines, and that the Euler characteristic computed from the CMB maps obtained by \emph{Planck}'s predecessor Wilkinson Microwave Anisotropy Probe (WMAP) satellite also exhibits mildly significant deviations between observations and simulations \citep{eriksen04ng}. Disregarding systematics, a more interesting possibility is the source of the anomaly being a genuine astrophysical signal, perhaps truly primordial in nature, or due to a yet unresolved foreground effect. If the signals are primordial, it opens up the possibility of the Universe admitting a  non-trivial global topology \citep{steinerhyperbolic,roukema2004,divalentino2019}, possibly induced by  large-scale topological defects \citep{durrer1999,vilenkin2000,bouchet2001,gangui2002}, as well as that of primordial non-Gaussianity \citep{verde2001,planckcollaboration2016a}. The latter scenario has justification based on the nature of the method used to detect anomalies. Methods based on comparing the power spectrum of different patches of the sky, as example hemispheres, inform about homogeneity properties \citep{eriksen2004,rst}, where as investigating the alignment of multipoles informs about isotropy properties \citep{multipoles}. Results based on these methods do not encode information about higher orders, and hence cannot shed light on non-Gaussianities. Computing higher order correlations is expensive, however $3$
-point correlation functions show the observations and simulations to be consistent (c.f. \cite{planckIsotropy2015,planckIsotropy2018}. Alternative methods, potentially encoding information of all orders emerge from geometry and topology. Principal such tools from integral geometry are the Minkowski functionals \citep{adler1981,mecke94,matsubara2010,ducout2013}, that also show consistent behavior between simulations and observations of CMB \citep{planckIsotropy2015,planckIsotropy2018}. In this scenario, the purely topological tools employed in this paper present, for the first time, an anomaly that may have contributing  influence from higher orders of correlation, thereby potentially pointing to a non-Gaussian signal. Future research will involve a coordinated effort in all the aforementioned directions to establish a deeper understanding of the origin and nature of the anomaly.

\section*{Acknowledgements}

I am greatly indebted to Robert Adler, Thomas Buchert, Herbert Edelsbrunner, Bernard Jones, Armin Schwarzman, Gert Vegter, and Rien van de Weygaert for encouraging this solo venture, and for acting as mentors over years. My gratitude also to Julian Borrill and Reijo Keskitalo for their patience in clarifying doubts, and their constructive comments on the draft. I also thank Tal Eliezri for insightful comments on the artwork. This work is supported by the ERC advanced grant ARThUs (grant no: 740021; PI: TB), with contributing influence from ERC advanced grant URSAT (grant no: 320422; PI: RA). I gratefully acknowledge the support of PSMN (P\^ole Scientifique de Mod\'elisation Num\'erique) of the ENS de Lyon, and the Department of Energy’s National Energy Research Scientific Computing Center (NERSC) at Lawrence Berkeley National Laboratory, operated under Contract No. DE-AC02-05CH11231, for the use of computing resources.


\bibliographystyle{aa}

\bibliography{/Users/pratyuze/papers/mine/references/master_references.bib}

\begin{thebibliography}{62}
\expandafter\ifx\csname natexlab\endcsname\relax\def\natexlab#1{#1}\fi

\bibitem[{Adler(1981)}]{adler1981}
Adler, R. 1981, The Geometry of Random Fields, Classics in applied mathematics
  (Society for Industrial and Applied Mathematics (SIAM, 3600 Market Street,
  Floor 6, Philadelphia, PA 19104))

\bibitem[{Adler {et~al.}(2017)Adler, Agami, \& Pranav}]{rst}
Adler, R.~J., Agami, S., \& Pranav, P. 2017, Proceedings of the National
  Academy of Sciences, 114, 11878

\bibitem[{{Appleby} {et~al.}(2020){Appleby}, {Park}, {Hong}, {Hwang}, \&
  {Kim}}]{appleby2020}
{Appleby}, S., {Park}, C., {Hong}, S.~E., {Hwang}, H.~S., \& {Kim}, J. 2020,
  \apj, 896, 145

\bibitem[{{Aurich} \& {Steiner}(2001)}]{steinerhyperbolic}
{Aurich}, R. \& {Steiner}, F. 2001, \mnras, 323, 1016

\bibitem[{{Bennett} {et~al.}(2013){Bennett}, {Larson}, {Weiland}, {Jarosik},
  {Hinshaw}, {Odegard}, {Smith}, {Hill}, {Gold}, {Halpern}, {Komatsu}, {Nolta},
  {Page}, {Spergel}, {Wollack}, {Dunkley}, {Kogut}, {Limon}, {Meyer}, {Tucker},
  \& {Wright}}]{wmap9}
{Bennett}, C.~L., {Larson}, D., {Weiland}, J.~L., {et~al.} 2013, Astrophys. J.
  Suppl., 208, 20

\bibitem[{{Biagetti} {et~al.}(2020){Biagetti}, {Cole}, \&
  {Shiu}}]{biagetti2020}
{Biagetti}, M., {Cole}, A., \& {Shiu}, G. 2020, arXiv e-prints,
  arXiv:2009.04819

\bibitem[{Bouchet {et~al.}(2001)Bouchet, Peter, Riazuelo, \&
  Sakellariadou}]{bouchet2001}
Bouchet, F.~R., Peter, P., Riazuelo, A., \& Sakellariadou, M. 2001, Phys. Rev.
  D, 65, 021301

\bibitem[{{Chingangbam} {et~al.}(2017){Chingangbam}, {Yogendran}, {Joby},
  {Ganesan}, {Appleby}, \& {Park}}]{chingangbam2017}
{Chingangbam}, P., {Yogendran}, K.~P., {Joby}, P.~K., {et~al.} 2017, \jcap, 12,
  023

\bibitem[{{Codis} {et~al.}(2013){Codis}, {Pichon}, {Pogosyan}, {Bernardeau}, \&
  {Matsubara}}]{codis2013}
{Codis}, S., {Pichon}, C., {Pogosyan}, D., {Bernardeau}, F., \& {Matsubara}, T.
  2013, \mnras, 435, 531

\bibitem[{Copi {et~al.}(2015)Copi, Huterer, Schwarz, \& Starkman}]{multipoles}
Copi, C., Huterer, D., Schwarz, D., \& Starkman, G. 2015, Monthly Notices of
  the Royal Astronomical Society, 449, 3458

\bibitem[{Di~Valentino {et~al.}(2019)Di~Valentino, Melchiorri, \&
  Silk}]{divalentino2019}
Di~Valentino, E., Melchiorri, A., \& Silk, J. 2019, Nature Astronomy

\bibitem[{{Ducout} {et~al.}(2013){Ducout}, {Bouchet}, {Colombi}, {Pogosyan}, \&
  {Prunet}}]{ducout2013}
{Ducout}, A., {Bouchet}, F.~R., {Colombi}, S., {Pogosyan}, D., \& {Prunet}, S.
  2013, MNRAS, 429, 2104

\bibitem[{{Durrer}(1999)}]{durrer1999}
{Durrer}, R. 1999, \nar, 43, 111

\bibitem[{Edelsbrunner \& Harer(2010)}]{edelsbrunnerharer10}
Edelsbrunner, H. \& Harer, J. 2010, Computational Topology - an Introduction
  (American Mathematical Society), I--XII, 1--241

\bibitem[{{Eriksen} {et~al.}(2004{\natexlab{a}}){Eriksen}, {Hansen}, {Banday},
  {G{\'o}rski}, \& {Lilje}}]{eriksen2004}
{Eriksen}, H.~K., {Hansen}, F.~K., {Banday}, A.~J., {G{\'o}rski}, K.~M., \&
  {Lilje}, P.~B. 2004{\natexlab{a}}, \apj, 605, 14

\bibitem[{{Eriksen} {et~al.}(2004{\natexlab{b}}){Eriksen}, {Novikov}, {Lilje},
  {Banday}, \& {G{\'o}rski}}]{eriksen04ng}
{Eriksen}, H.~K., {Novikov}, D.~I., {Lilje}, P.~B., {Banday}, A.~J., \&
  {G{\'o}rski}, K.~M. 2004{\natexlab{b}}, \apj, 612, 64

\bibitem[{Euler(1758)}]{euler1758}
Euler, L. 1758, Novi Commentarii academiae scientiarum Petropolitanae, 4, 140

\bibitem[{Feldbrugge {et~al.}(2019)Feldbrugge, van Engelen, van~de Weygaert,
  Pranav, \& Vegter}]{feldbrugge2019}
Feldbrugge, J., van Engelen, M., van~de Weygaert, R., Pranav, P., \& Vegter, G.
  2019, Journal of Cosmology and Astroparticle Physics, 2019, 052–052

\bibitem[{{Gangui}(2002)}]{gangui2002}
{Gangui}, A. 2002, International Journal of Modern Physics A, 17, 4273

\bibitem[{{Gauss}(1900)}]{gauss1900}
{Gauss}, C.~F. 1900, K. Gesellschaft Wissenschaft, 8

\bibitem[{{G{\'o}rski} {et~al.}(2005){G{\'o}rski}, {Hivon}, {Banday},
  {Wandelt}, {Hansen}, {Reinecke}, \& {Bartelmann}}]{healpix1}
{G{\'o}rski}, K.~M., {Hivon}, E., {Banday}, A.~J., {et~al.} 2005, Astrophysical
  Journal, 622, 759

\bibitem[{{Gott} {et~al.}(1986){Gott}, {Dickinson}, \& {Melott}}]{gdm86}
{Gott}, III, J.~R., {Dickinson}, M., \& {Melott}, A.~L. 1986, Astrophysical
  Journal, 306, 341

\bibitem[{{Guth}(1981)}]{guth1981}
{Guth}, A.~H. 1981, Physical Review D, 23, 347

\bibitem[{{Guth} \& {Pi}(1982)}]{guthpi1982}
{Guth}, A.~H. \& {Pi}, S.-Y. 1982, Physical Review Letters, 49, 1110

\bibitem[{{Harrison}(1970)}]{harrison1970}
{Harrison}, E.~R. 1970, \prd, 1, 2726

\bibitem[{Jaffe {et~al.}(2001)Jaffe, Ade, Balbi, Bock, Bond, Borrill,
  Boscaleri, Coble, Crill, de~Bernardis, Farese, Ferreira, Ganga, Giacometti,
  Hanany, Hivon, Hristov, Iacoangeli, Lange, Lee, Martinis, Masi, Mauskopf,
  Melchiorri, Montroy, Netterfield, Oh, Pascale, Piacentini, Pogosyan, Prunet,
  Rabii, Rao, Richards, Romeo, Ruhl, Scaramuzzi, Sforna, Smoot, Stompor,
  Winant, \& Wu}]{jaffe2001}
Jaffe, A.~H., Ade, P. A.~R., Balbi, A., {et~al.} 2001, Phys. Rev. Lett., 86,
  3475

\bibitem[{{Jones}(2017)}]{jones2017precision}
{Jones}, B.~J.~T. 2017, {Precision Cosmology: The First Half Million Years}
  (Cambridge University Press)

\bibitem[{Kono {et~al.}(2020)Kono, Takeuchi, Cooray, Nishizawa, \&
  Murakami}]{kono2020}
Kono, K.~T., Takeuchi, T.~T., Cooray, S., Nishizawa, A.~J., \& Murakami, K.
  2020, arXiv e-prints, arXiv:2006.02905

\bibitem[{Mahalanobis(1936)}]{mahalanobis}
Mahalanobis, P.~C. 1936, in Proceedings National Institute of Science, India,
  Vol.~2, 49--55

\bibitem[{{Masi}(2002)}]{boomerang}
{Masi}, S. 2002, Progress in Particle and Nuclear Physics, 48, 243

\bibitem[{{Matsubara}(2010)}]{matsubara2010}
{Matsubara}, T. 2010, \prd, 81, 083505

\bibitem[{{Mecke} {et~al.}(1994){Mecke}, {Buchert}, \& {Wagner}}]{mecke94}
{Mecke}, K.~R., {Buchert}, T., \& {Wagner}, H. 1994, Astronomy \& Astrophysics,
  288, 697

\bibitem[{Munkres(1984)}]{munkres1984}
Munkres, J. 1984, Elements of Algebraic Topology, Advanced book classics
  (Perseus Books)

\bibitem[{{Park} {et~al.}(2013){Park}, {Pranav}, {Chingangbam}, {van de
  Weygaert}, {Jones}, {Vegter}, {Kim}, {Hidding}, \& {Hellwing}}]{ppc13}
{Park}, C., {Pranav}, P., {Chingangbam}, P., {et~al.} 2013, Journal of Korean
  Astronomical Society, 46, 125

\bibitem[{{Peebles} \& {Yu}(1970)}]{peebles1970}
{Peebles}, P.~J.~E. \& {Yu}, J.~T. 1970, \apj, 162, 815

\bibitem[{{Planck Collaboration} {et~al.}(2016{\natexlab{a}}){Planck
  Collaboration}, {Ade}, {Aghanim}, {Akrami}, {Aluri}, {Arnaud}, {Ashdown},
  {Aumont}, {Baccigalupi}, {Banday}, {Barreiro}, {Bartolo}, {Basak},
  {Battaner}, {Benabed}, {Beno{\^\i}t}, {Benoit-L{\'e}vy}, {Bernard},
  {Bersanelli}, {Bielewicz}, {Bock}, {Bonaldi}, {Bonavera}, {Bond}, {Borrill},
  {Bouchet}, {Boulanger}, {Bucher}, {Burigana}, {Butler}, {Calabrese},
  {Cardoso}, {Casaponsa}, {Catalano}, {Challinor}, {Chamballu}, {Chiang},
  {Christensen}, {Church}, {Clements}, {Colombi}, {Colombo}, {Combet},
  {Contreras}, {Couchot}, {Coulais}, {Crill}, {Cruz}, {Curto}, {Cuttaia},
  {Danese}, {Davies}, {Davis}, {de Bernardis}, {de Rosa}, {de Zotti},
  {Delabrouille}, {D{\'e}sert}, {Diego}, {Dole}, {Donzelli}, {Dor{\'e}},
  {Douspis}, {Ducout}, {Dupac}, {Efstathiou}, {Elsner}, {En{\ss}lin},
  {Eriksen}, {Fantaye}, {Fergusson}, {Fernandez-Cobos}, {Finelli}, {Forni},
  {Frailis}, {Fraisse}, {Franceschi}, {Frejsel}, {Frolov}, {Galeotta}, {Galli},
  {Ganga}, {Gauthier}, {Ghosh}, {Giard}, {Giraud-H{\'e}raud}, {Gjerl{\o}w},
  {Gonz{\'a}lez-Nuevo}, {G{\'o}rski}, {Gratton}, {Gregorio}, {Gruppuso},
  {Gudmundsson}, {Hansen}, {Hanson}, {Harrison}, {Henrot-Versill{\'e}},
  {Hern{\'a}ndez-Monteagudo}, {Herranz}, {Hildebrandt}, {Hivon}, {Hobson},
  {Holmes}, {Hornstrup}, {Hovest}, {Huang}, {Huffenberger}, {Hurier}, {Jaffe},
  {Jaffe}, {Jones}, {Juvela}, {Keih{\"a}nen}, {Keskitalo}, {Kim}, {Kisner},
  {Knoche}, {Kunz}, {Kurki-Suonio}, {Lagache}, {L{\"a}hteenm{\"a}ki},
  {Lamarre}, {Lasenby}, {Lattanzi}, {Lawrence}, {Leonardi}, {Lesgourgues},
  {Levrier}, {Liguori}, {Lilje}, {Linden-V{\o}rnle}, {Liu},
  {L{\'o}pez-Caniego}, {Lubin}, {Mac{\'\i}as-P{\'e}rez}, {Maggio}, {Maino},
  {Mandolesi}, {Mangilli}, {Marinucci}, {Maris}, {Martin},
  {Mart{\'\i}nez-Gonz{\'a}lez}, {Masi}, {Matarrese}, {McGehee}, {Meinhold},
  {Melchiorri}, {Mendes}, {Mennella}, {Migliaccio}, {Mikkelsen}, {Mitra},
  {Miville-Desch{\^e}nes}, {Molinari}, {Moneti}, {Montier}, {Morgante},
  {Mortlock}, {Moss}, {Munshi}, {Murphy}, {Naselsky}, {Nati}, {Natoli},
  {Netterfield}, {N{\o}rgaard-Nielsen}, {Noviello}, {Novikov}, {Novikov},
  {Oxborrow}, {Paci}, {Pagano}, {Pajot}, \& et.al.}]{planckIsotropy2015}
{Planck Collaboration}, {Ade}, P.~A.~R., {Aghanim}, N., {et~al.}
  2016{\natexlab{a}}, \aap, 594, A16

\bibitem[{{Planck Collaboration} {et~al.}(2016{\natexlab{b}}){Planck
  Collaboration}, {Ade}, {Aghanim}, {Arnaud}, {Arroja}, {Ashdown}, {Aumont},
  {Baccigalupi}, {Ballardini}, {Banday}, \& et~al.}]{planckcollaboration2016a}
{Planck Collaboration}, {Ade}, P.~A.~R., {Aghanim}, N., {et~al.}
  2016{\natexlab{b}}, \aap, 594, A17

\bibitem[{{Planck Collaboration} {et~al.}(2016{\natexlab{c}}){Planck
  Collaboration}, {Ade}, {Aghanim}, {Arnaud}, {Ashdown}, {Aumont},
  {Baccigalupi}, {Banday}, {Barreiro}, {Bartlett}, \& et~al.}]{plancksims}
{Planck Collaboration}, {Ade}, P.~A.~R., {Aghanim}, N., {et~al.}
  2016{\natexlab{c}}, \aap, 594, A12

\bibitem[{{Planck Collaboration} {et~al.}(2020{\natexlab{a}}){Planck
  Collaboration}, {Aghanim}, {Akrami}, {Arroja}, {Ashdown}, {Aumont},
  {Baccigalupi}, {Ballardini}, {Banday}, {Barreiro}, {Bartolo}, {Basak},
  {Battye}, {Benabed}, {Bernard}, {Bersanelli}, {Bielewicz}, {Bock}, {Bond},
  {Borrill}, {Bouchet}, {Boulanger}, {Bucher}, {Burigana}, {Butler},
  {Calabrese}, {Cardoso}, {Carron}, {Casaponsa}, {Challinor}, {Chiang},
  {Colombo}, {Combet}, {Contreras}, {Crill}, {Cuttaia}, {de Bernardis}, {de
  Zotti}, {Delabrouille}, {Delouis}, {D{\'e}sert}, {Di Valentino}, {Dickinson},
  {Diego}, {Donzelli}, {Dor{\'e}}, {Douspis}, {Ducout}, {Dupac}, {Efstathiou},
  {Elsner}, {En{\ss}lin}, {Eriksen}, {Falgarone}, {Fantaye}, {Fergusson},
  {Fernandez-Cobos}, {Finelli}, {Forastieri}, {Frailis}, {Franceschi},
  {Frolov}, {Galeotta}, {Galli}, {Ganga}, {G{\'e}nova-Santos}, {Gerbino},
  {Ghosh}, {Gonz{\'a}lez-Nuevo}, {G{\'o}rski}, {Gratton}, {Gruppuso},
  {Gudmundsson}, {Hamann}, {Handley}, {Hansen}, {Helou}, {Herranz},
  {Hildebrandt}, {Hivon}, {Huang}, {Jaffe}, {Jones}, {Karakci}, {Keih{\"a}nen},
  {Keskitalo}, {Kiiveri}, {Kim}, {Kisner}, {Knox}, {Krachmalnicoff}, {Kunz},
  {Kurki-Suonio}, {Lagache}, {Lamarre}, {Langer}, {Lasenby}, {Lattanzi},
  {Lawrence}, {Le Jeune}, {Leahy}, {Lesgourgues}, {Levrier}, {Lewis},
  {Liguori}, {Lilje}, {Lilley}, {Lindholm}, {L{\'o}pez-Caniego}, {Lubin}, {Ma},
  {Mac{\'\i}as-P{\'e}rez}, {Maggio}, {Maino}, {Mandolesi}, {Mangilli},
  {Marcos-Caballero}, {Maris}, {Martin}, {Martinelli},
  {Mart{\'\i}nez-Gonz{\'a}lez}, {Matarrese}, {Mauri}, {McEwen}, {Meerburg},
  {Meinhold}, {Melchiorri}, {Mennella}, {Migliaccio}, {Millea}, {Mitra},
  {Miville-Desch{\^e}nes}, {Molinari}, {Moneti}, {Montier}, {Morgante}, {Moss},
  {Mottet}, {M{\"u}nchmeyer}, {Natoli}, {N{\o}rgaard-Nielsen}, {Oxborrow},
  {Pagano}, {Paoletti}, {Partridge}, {Patanchon}, {Pearson}, {Peel}, {Peiris},
  {Perrotta}, {Pettorino}, {Piacentini}, {Polastri}, {Polenta}, {Puget},
  {Rachen}, {Reinecke}, {Remazeilles}, {Renault}, {Renzi}, {Rocha}, {Rosset},
  {Roudier}, {Rubi{\~n}o-Mart{\'\i}n}, {Ruiz-Granados}, {Salvati}, {Sandri},
  {Savelainen}, {Scott}, {Shellard}, {Shiraishi}, {Sirignano}, {Sirri},
  {Spencer}, {Sunyaev}, {Suur-Uski}, {Tauber}, {Tavagnacco}, {Tenti},
  {Terenzi}, {Toffolatti}, {Tomasi}, {Trombetti}, {Valiviita}, {Van Tent},
  {Vibert}, {Vielva}, {Villa}, {Vittorio}, {Wandelt}, {Wehus}, {White},
  {White}, {Zacchei}, \& {Zonca}}]{planckOverview2018}
{Planck Collaboration}, {Aghanim}, N., {Akrami}, Y., {et~al.}
  2020{\natexlab{a}}, \aap, 641, A1

\bibitem[{{Planck Collaboration} {et~al.}(2020{\natexlab{b}}){Planck
  Collaboration}, {Akrami}, {Andersen}, {Ashdown}, {Baccigalupi}, {Ballardini},
  {Banday}, {Barreiro}, {Bartolo}, {Basak}, {Benabed}, {Bernard}, {Bersanelli},
  {Bielewicz}, {Bond}, {Borrill}, {Burigana}, {Butler}, {Calabrese},
  {Casaponsa}, {Chiang}, {Colombo}, {Combet}, {Crill}, {Cuttaia}, {de
  Bernardis}, {de Rosa}, {de Zotti}, {Delabrouille}, {Di Valentino}, {Diego},
  {Dor{\'e}}, {Douspis}, {Dupac}, {Eriksen}, {Fernandez-Cobos}, {Finelli},
  {Frailis}, {Fraisse}, {Franceschi}, {Frolov}, {Galeotta}, {Galli}, {Ganga},
  {Gerbino}, {Ghosh}, {Gonz{\'a}lez-Nuevo}, {G{\'o}rski}, {Gruppuso},
  {Gudmundsson}, {Handley}, {Helou}, {Herranz}, {Hildebrandt}, {Hivon},
  {Huang}, {Jaffe}, {Jones}, {Keih{\"a}nen}, {Keskitalo}, {Kiiveri}, {Kim},
  {Kisner}, {Krachmalnicoff}, {Kunz}, {Kurki-Suonio}, {Lasenby}, {Lattanzi},
  {Lawrence}, {Le Jeune}, {Levrier}, {Liguori}, {Lilje}, {Lilley}, {Lindholm},
  {L{\'o}pez-Caniego}, {Lubin}, {Mac{\'\i}as-P{\'e}rez}, {Maino}, {Mandolesi},
  {Marcos-Caballero}, {Maris}, {Martin}, {Mart{\'\i}nez-Gonz{\'a}lez},
  {Matarrese}, {Mauri}, {McEwen}, {Meinhold}, {Mennella}, {Migliaccio},
  {Mitra}, {Molinari}, {Montier}, {Morgante}, {Moss}, {Natoli}, {Paoletti},
  {Partridge}, {Patanchon}, {Pearson}, {Pearson}, {Perrotta}, {Piacentini},
  {Polenta}, {Rachen}, {Reinecke}, {Remazeilles}, {Renzi}, {Rocha}, {Rosset},
  {Roudier}, {Rubi{\~n}o-Mart{\'\i}n}, {Ruiz-Granados}, {Salvati},
  {Savelainen}, {Scott}, {Sirignano}, {Sirri}, {Spencer}, {Suur-Uski},
  {Svalheim}, {Tauber}, {Tavagnacco}, {Tenti}, {Terenzi}, {Thommesen},
  {Toffolatti}, {Tomasi}, {Tristram}, {Trombetti}, {Valiviita}, {Van Tent},
  {Vielva}, {Villa}, {Vittorio}, {Wandelt}, {Wehus}, {Zacchei}, \&
  {Zonca}}]{npipe}
{Planck Collaboration}, {Akrami}, Y., {Andersen}, K.~J., {et~al.}
  2020{\natexlab{b}}, \aap, 643, A42

\bibitem[{{Planck Collaboration} {et~al.}(2020{\natexlab{c}}){Planck
  Collaboration}, {Akrami}, {Ashdown}, {Aumont}, {Baccigalupi}, {Ballardini},
  {Banday}, {Barreiro}, {Bartolo}, {Basak}, {Benabed}, {Bersanelli},
  {Bielewicz}, {Bock}, {Bond}, {Borrill}, {Bouchet}, {Boulanger}, {Bucher},
  {Burigana}, {Butler}, {Calabrese}, {Cardoso}, {Casaponsa}, {Chiang},
  {Colombo}, {Combet}, {Contreras}, {Crill}, {de Bernardis}, {de Zotti},
  {Delabrouille}, {Delouis}, {Di Valentino}, {Diego}, {Dor{\'e}}, {Douspis},
  {Ducout}, {Dupac}, {Efstathiou}, {Elsner}, {En{\ss}lin}, {Eriksen},
  {Fantaye}, {Fernandez-Cobos}, {Finelli}, {Frailis}, {Fraisse}, {Franceschi},
  {Frolov}, {Galeotta}, {Galli}, {Ganga}, {G{\'e}nova-Santos}, {Gerbino},
  {Ghosh}, {Gonz{\'a}lez-Nuevo}, {G{\'o}rski}, {Gruppuso}, {Gudmundsson},
  {Hamann}, {Handley}, {Hansen}, {Herranz}, {Hivon}, {Huang}, {Jaffe}, {Jones},
  {Keih{\"a}nen}, {Keskitalo}, {Kiiveri}, {Kim}, {Krachmalnicoff}, {Kunz},
  {Kurki-Suonio}, {Lagache}, {Lamarre}, {Lasenby}, {Lattanzi}, {Lawrence}, {Le
  Jeune}, {Levrier}, {Liguori}, {Lilje}, {Lindholm}, {L{\'o}pez-Caniego}, {Ma},
  {Mac{\'\i}as-P{\'e}rez}, {Maggio}, {Maino}, {Mandolesi}, {Mangilli},
  {Marcos-Caballero}, {Maris}, {Martin}, {Mart{\'\i}nez-Gonz{\'a}lez},
  {Matarrese}, {Mauri}, {McEwen}, {Meinhold}, {Mennella}, {Migliaccio},
  {Miville-Desch{\^e}nes}, {Molinari}, {Moneti}, {Montier}, {Morgante}, {Moss},
  {Natoli}, {Pagano}, {Paoletti}, {Partridge}, {Perrotta}, {Pettorino},
  {Piacentini}, {Polenta}, {Puget}, {Rachen}, {Reinecke}, {Remazeilles},
  {Renzi}, {Rocha}, {Rosset}, {Roudier}, {Rubi{\~n}o-Mart{\'\i}n},
  {Ruiz-Granados}, {Salvati}, {Savelainen}, {Scott}, {Shellard}, {Sirignano},
  {Sunyaev}, {Suur-Uski}, {Tauber}, {Tavagnacco}, {Tenti}, {Toffolatti},
  {Tomasi}, {Trombetti}, {Valenziano}, {Valiviita}, {Van Tent}, {Vielva},
  {Villa}, {Vittorio}, {Wandelt}, {Wehus}, {Zacchei}, {Zibin}, \&
  {Zonca}}]{planckIsotropy2018}
{Planck Collaboration}, {Akrami}, Y., {Ashdown}, M., {et~al.}
  2020{\natexlab{c}}, \aap, 641, A7

\bibitem[{{Planck Collaboration} {et~al.}(2020{\natexlab{d}}){Planck
  Collaboration}, {Akrami}, {Ashdown}, {Aumont}, {Baccigalupi}, {Ballardini},
  {Banday}, {Barreiro}, {Bartolo}, {Basak}, {Benabed}, {Bersanelli},
  {Bielewicz}, {Bond}, {Borrill}, {Bouchet}, {Boulanger}, {Bucher}, {Burigana},
  {Calabrese}, {Cardoso}, {Carron}, {Casaponsa}, {Challinor}, {Colombo},
  {Combet}, {Crill}, {Cuttaia}, {de Bernardis}, {de Rosa}, {de Zotti},
  {Delabrouille}, {Delouis}, {Di Valentino}, {Dickinson}, {Diego}, {Donzelli},
  {Dor{\'e}}, {Ducout}, {Dupac}, {Efstathiou}, {Elsner}, {En{\ss}lin},
  {Eriksen}, {Falgarone}, {Fernandez-Cobos}, {Finelli}, {Forastieri},
  {Frailis}, {Fraisse}, {Franceschi}, {Frolov}, {Galeotta}, {Galli}, {Ganga},
  {G{\'e}nova-Santos}, {Gerbino}, {Ghosh}, {Gonz{\'a}lez-Nuevo}, {G{\'o}rski},
  {Gratton}, {Gruppuso}, {Gudmundsson}, {Handley}, {Hansen}, {Helou},
  {Herranz}, {Hildebrandt}, {Huang}, {Jaffe}, {Karakci}, {Keih{\"a}nen},
  {Keskitalo}, {Kiiveri}, {Kim}, {Kisner}, {Krachmalnicoff}, {Kunz},
  {Kurki-Suonio}, {Lagache}, {Lamarre}, {Lasenby}, {Lattanzi}, {Lawrence}, {Le
  Jeune}, {Levrier}, {Liguori}, {Lilje}, {Lindholm}, {L{\'o}pez-Caniego},
  {Lubin}, {Ma}, {Mac{\'\i}as-P{\'e}rez}, {Maggio}, {Maino}, {Mandolesi},
  {Mangilli}, {Marcos-Caballero}, {Maris}, {Martin},
  {Mart{\'\i}nez-Gonz{\'a}lez}, {Matarrese}, {Mauri}, {McEwen}, {Meinhold},
  {Melchiorri}, {Mennella}, {Migliaccio}, {Miville-Desch{\^e}nes}, {Molinari},
  {Moneti}, {Montier}, {Morgante}, {Natoli}, {Oppizzi}, {Pagano}, {Paoletti},
  {Partridge}, {Peel}, {Pettorino}, {Piacentini}, {Polenta}, {Puget}, {Rachen},
  {Reinecke}, {Remazeilles}, {Renzi}, {Rocha}, {Roudier},
  {Rubi{\~n}o-Mart{\'\i}n}, {Ruiz-Granados}, {Salvati}, {Sandri}, {Savelainen},
  {Scott}, {Seljebotn}, {Sirignano}, {Spencer}, {Suur-Uski}, {Tauber},
  {Tavagnacco}, {Tenti}, {Thommesen}, {Toffolatti}, {Tomasi}, {Trombetti},
  {Valiviita}, {Van Tent}, {Vielva}, {Villa}, {Vittorio}, {Wandelt}, {Wehus},
  {Zacchei}, \& {Zonca}}]{ffp10}
{Planck Collaboration}, {Akrami}, Y., {Ashdown}, M., {et~al.}
  2020{\natexlab{d}}, \aap, 641, A4

\bibitem[{{Pogosyan} {et~al.}(2009){Pogosyan}, {Gay}, \&
  {Pichon}}]{pogosyan2009}
{Pogosyan}, D., {Gay}, C., \& {Pichon}, C. 2009, \prd, 80, 081301

\bibitem[{Pranav(2015)}]{pranavthesis}
Pranav, P. 2015, Persistent Holes in the Universe: A Hierarchical Topology of
  the Cosmic Mass Distribution (University of Groningen)

\bibitem[{{Pranav} {et~al.}(2019{\natexlab{a}}){Pranav}, {Adler}, {Buchert},
  {Edelsbrunner}, {Jones}, {Schwartzman}, {Wagner}, \& {van de
  Weygaert}}]{pranav2019b}
{Pranav}, P., {Adler}, R.~J., {Buchert}, T., {et~al.} 2019{\natexlab{a}}, \aap,
  627, A163

\bibitem[{{Pranav} {et~al.}(2017){Pranav}, {Edelsbrunner}, {van de Weygaert},
  {Vegter}, {Kerber}, {Jones}, \& {Wintraecken}}]{pranav2017}
{Pranav}, P., {Edelsbrunner}, H., {van de Weygaert}, R., {et~al.} 2017, \mnras,
  465, 4281

\bibitem[{{Pranav} {et~al.}(2019{\natexlab{b}}){Pranav}, {van de Weygaert},
  {Vegter}, {Jones}, {Adler}, {Feldbrugge}, {Park}, {Buchert}, \&
  {Kerber}}]{pranav2019a}
{Pranav}, P., {van de Weygaert}, R., {Vegter}, G., {et~al.} 2019{\natexlab{b}},
  \mnras, 485, 4167

\bibitem[{{Roukema} {et~al.}(2004){Roukema}, {Lew}, {Cechowska}, {Marecki}, \&
  {Bajtlik}}]{roukema2004}
{Roukema}, B.~F., {Lew}, B., {Cechowska}, M., {Marecki}, A., \& {Bajtlik}, S.
  2004, \aap, 423, 821

\bibitem[{Ryden(2003)}]{ryden2003}
Ryden, B. 2003, Introduction to Cosmology (Addison-Wesley)

\bibitem[{Sahni {et~al.}(1998)Sahni, Sathyprakash, \& Shandarin}]{sahni1998}
Sahni, V., Sathyprakash, B., \& Shandarin, S. 1998, {\apj}, 507, L109

\bibitem[{{Schmalzing} \& {Gorski}(1998)}]{schmalzinggorski}
{Schmalzing}, J. \& {Gorski}, K.~M. 1998, \mnras, 297, 355

\bibitem[{{Schwarz} {et~al.}(2016){Schwarz}, {Copi}, {Huterer}, \&
  {Starkman}}]{cmbanomaliesstarkman}
{Schwarz}, D.~J., {Copi}, C.~J., {Huterer}, D., \& {Starkman}, G.~D. 2016,
  Classical and Quantum Gravity, 33, 184001

\bibitem[{Shivashankar {et~al.}(2016)Shivashankar, Pranav, Natarajan, van~de
  Weygaert, Bos, \& Rieder}]{shivashankar2015}
Shivashankar, N., Pranav, P., Natarajan, V., {et~al.} 2016, {IEEE} Trans. Vis.
  Comput. Graph., 22, 1745

\bibitem[{{Starobinsky}(1982)}]{starobinsky1982}
{Starobinsky}, A.~A. 1982, Physics Letters B, 117, 175

\bibitem[{{Telschow} {et~al.}(2019){Telschow}, {Schwartzman}, {Cheng}, \&
  {Pranav}}]{eecestimate}
{Telschow}, F., {Schwartzman}, A., {Cheng}, D., \& {Pranav}, P. 2019, arXiv
  e-prints, arXiv:1908.02493

\bibitem[{Tukey(1975)}]{depth}
Tukey, J.~W. 1975, in Proceedings of the 1974 international congress of
  mathematicians, Vol.~2, 523--531

\bibitem[{{van de Weygaert} {et~al.}(2011){van de Weygaert}, {Pranav}, {Jones},
  {Bos}, {Vegter}, {Edelsbrunner}, {Teillaud}, {Hellwing}, {Park}, {Hidding},
  \& {Wintraecken}}]{pranavde}
{van de Weygaert}, R., {Pranav}, P., {Jones}, B.~J.~T., {et~al.} 2011, ArXiv
  e-prints, arXiv:1110.5528

\bibitem[{van~de Weygaert {et~al.}(2011)van~de Weygaert, Vegter, Edelsbrunner,
  Jones, Pranav, Park, Hellwing, Eldering, Kruithof, Bos, Hidding, Feldbrugge,
  ten Have, van Engelen, Caroli, \& Teillaud}]{isvd10}
van~de Weygaert, R., Vegter, G., Edelsbrunner, H., {et~al.} 2011, Transactions
  on Computational Science, 14, 60

\bibitem[{{Verde} {et~al.}(2001){Verde}, {Jimenez}, {Kamionkowski}, \&
  {Matarrese}}]{verde2001}
{Verde}, L., {Jimenez}, R., {Kamionkowski}, M., \& {Matarrese}, S. 2001,
  \mnras, 325, 412

\bibitem[{{Vilenkin} \& {Shellard}(2000)}]{vilenkin2000}
{Vilenkin}, A. \& {Shellard}, E.~P.~S. 2000, {Cosmic Strings and Other
  Topological Defects} (Cambridge University Press)

\bibitem[{{Wilding} {et~al.}(2020){Wilding}, {Nevenzeel}, {van de Weygaert},
  {Vegter}, {Pranav}, {Jones}, {Efstathiou}, \& {Feldbrugge}}]{wilding2020}
{Wilding}, G., {Nevenzeel}, K., {van de Weygaert}, R., {et~al.} 2020, arXiv
  e-prints, arXiv:2011.12851

\bibitem[{Xu {et~al.}(2019)Xu, Cisewski-Kehe, Green, \& Nagai}]{xu2019}
Xu, X., Cisewski-Kehe, J., Green, S., \& Nagai, D. 2019, Astronomy and
  Computing, 27, 34–52

\end{thebibliography}

\begin{appendix}
\appendix

\section{Data and methods}
\label{sec:methods}

In this Section, we briefly describe the dataset used in the experiments. We also present a short account of the computational pipeline, referring the reader to \cite{pranav2019b} for details.

\subsection{Datasets}

Over a period of time, the Planck team has invested significant effort in understanding and calibrating the source of noise as well as systematics that affect observations. This has led to the release of a series of datasets over a period of time, with successive data releases achieving better calibration and accuracy. In \cite{pranav2019b}, we performed our experiments on the second data release (DR2 hereafter). which is one of the intermediate data releases. In this paper, we concentrate on two data releases simultaneously, which we briefly describe below. 

\medskip\noindent\textbf{\emph{Planck 2020} Data Release 4 \texttt{NPIPE} dataset.} The primary results are based on the observational maps and simulations based on the \texttt{NPIPE} \citep{npipe} analysis pipeline (hereafter just \texttt{NPIPE}), which employs the \texttt{SEVEM} component separation technique. The \texttt{NPIPE} dataset is the final data release from the Planck team, and the pipeline is the most evolved and sophisticated of all  data generation pipelines. It combines some of the most powerful features of the separate LFI and HFI analysis piplelines, resulting in frequency and component maps that have lower levels of noise and systematics at essentially all angular scales (c.f. \cite{npipe}). The observational maps are accompanied by a suite of $600$ simulations modeled as isotropic, homogeneous Gaussian random fields.

\medskip\noindent\textbf{\emph{Planck 2018} Data Release 3 \texttt{FFP10} dataset.} For comparison, we also analyze the temperature maps from \emph{Planck 2018} full-mission data release (\texttt{DR3} hereafter), which is the third data release by the Planck team, and employs the \texttt{SMICA} component separation technique. The noise and systematics in this release are better constrained than in the previous data releases. The component separated observational maps are accompanied by a  suite of  $300$ simulations generated using the \emph{Full Focal Plane Plane} pipeline (hereafter referred to as the \texttt{FFP10} simulations) \citep{ffp10}, also based on the standard model.

\subsection{Computational pipeline}

\begin{figure}
	\centering
	\subfloat{\includegraphics[width=0.7\textwidth]{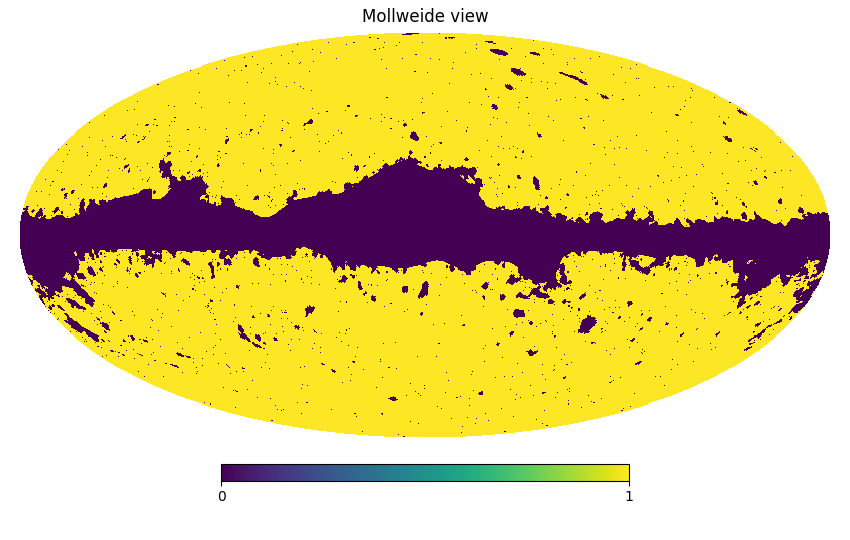} }\\
	\caption{A visualization of the mask employed for the analyses in this paper. It is the common mask released in Data release 3 (2018), and masks our galaxy as well as other bright foreground sources, including point sources.}
	\label{fig:mask}
\end{figure}

\begin{figure*}
	\centering
	\subfloat{\includegraphics[width=0.4\textwidth]{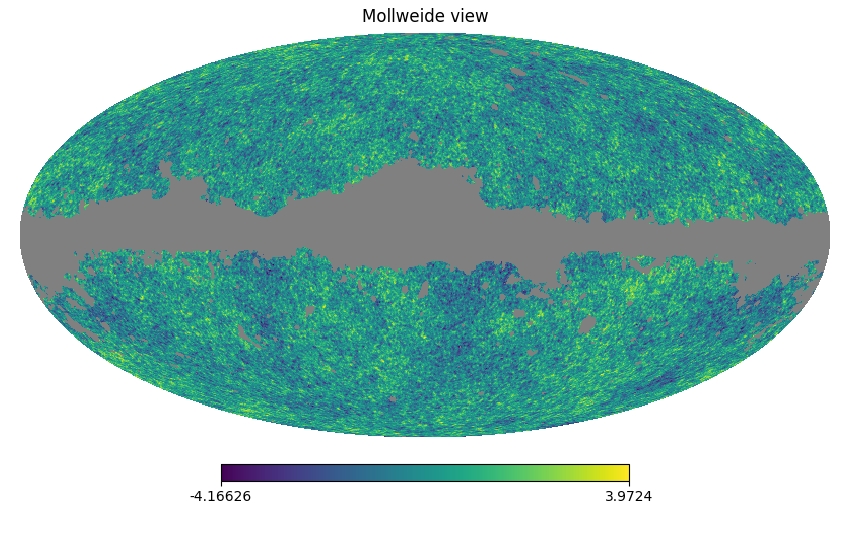} }
	\subfloat{\includegraphics[width=0.4\textwidth]{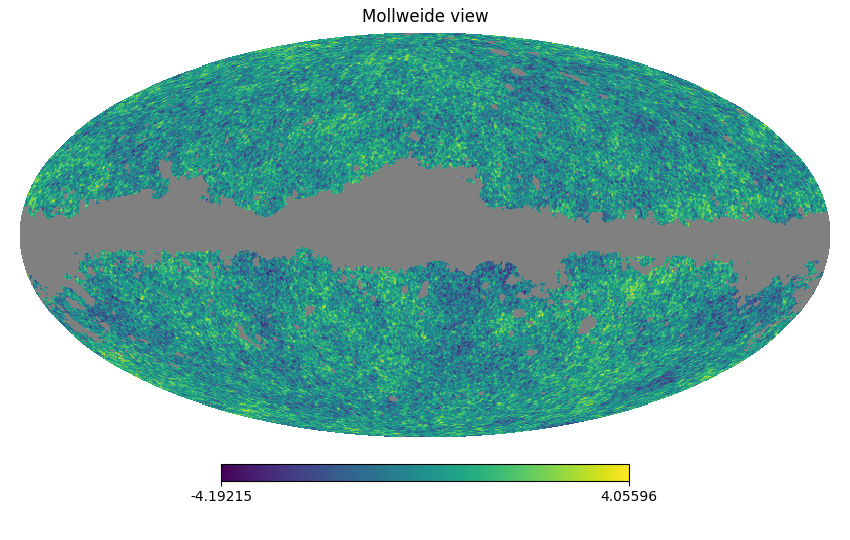} }\\
	\subfloat{\includegraphics[width=0.4\textwidth]{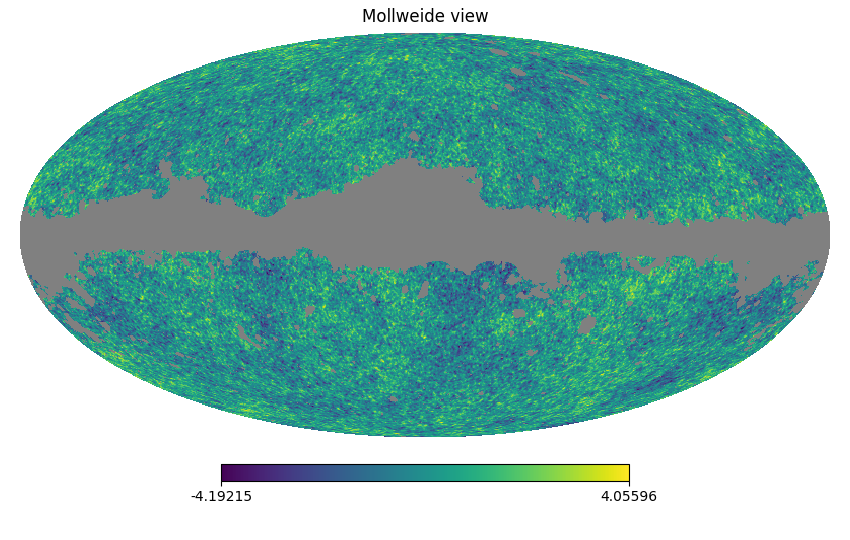} }
	\subfloat{\includegraphics[width=0.4\textwidth]{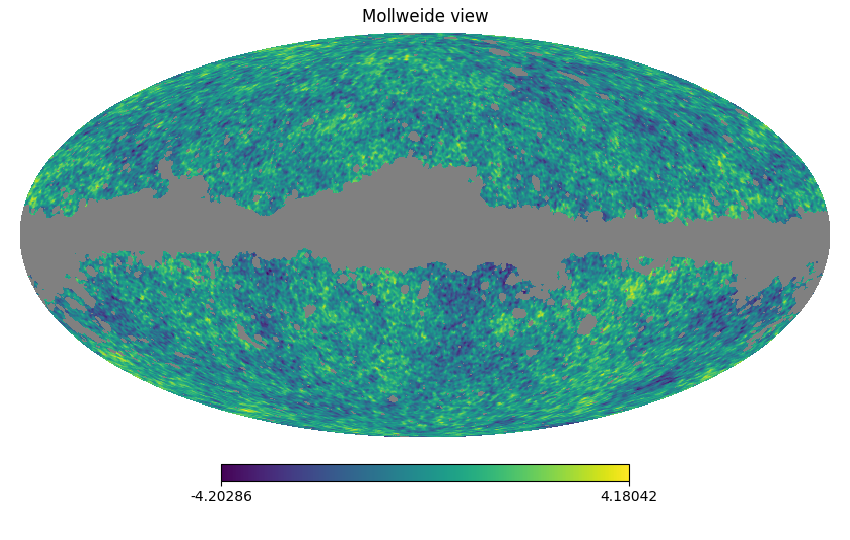} }\\
	\subfloat{\includegraphics[width=0.4\textwidth]{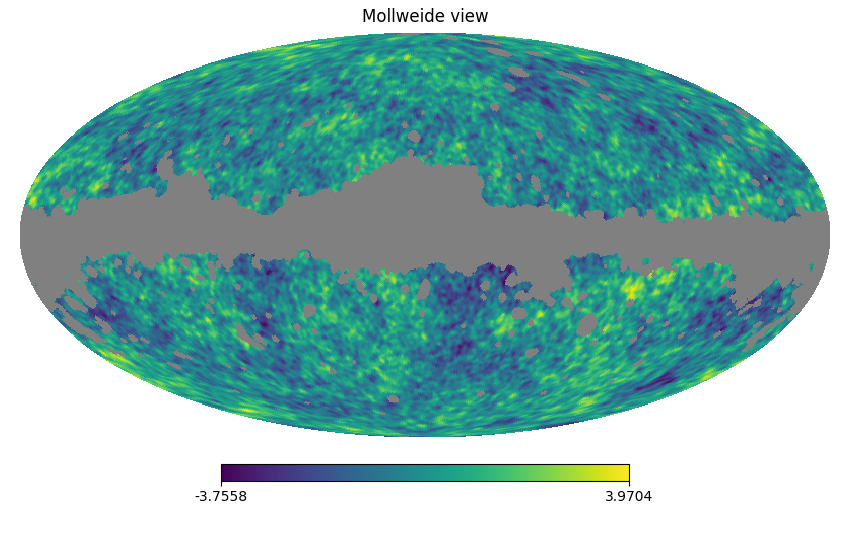} }
	\subfloat{\includegraphics[width=0.4\textwidth]{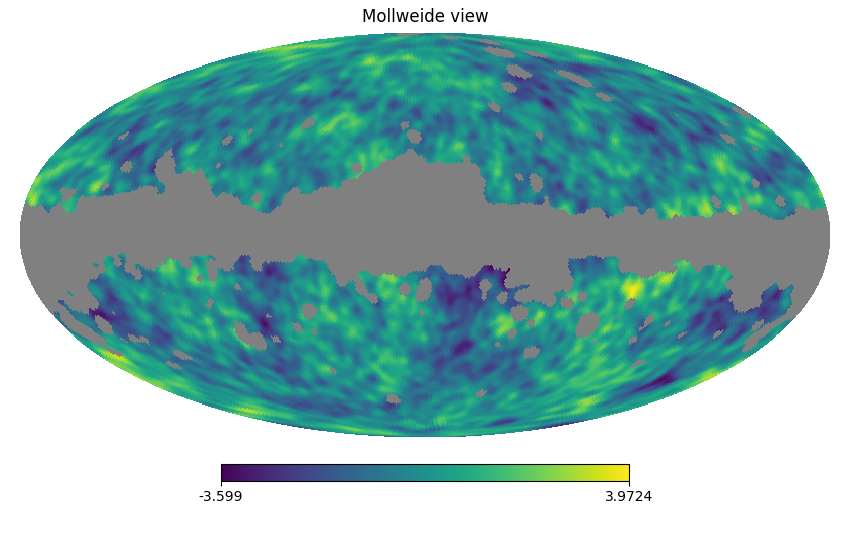} }\\
	\subfloat{\includegraphics[width=0.4\textwidth]{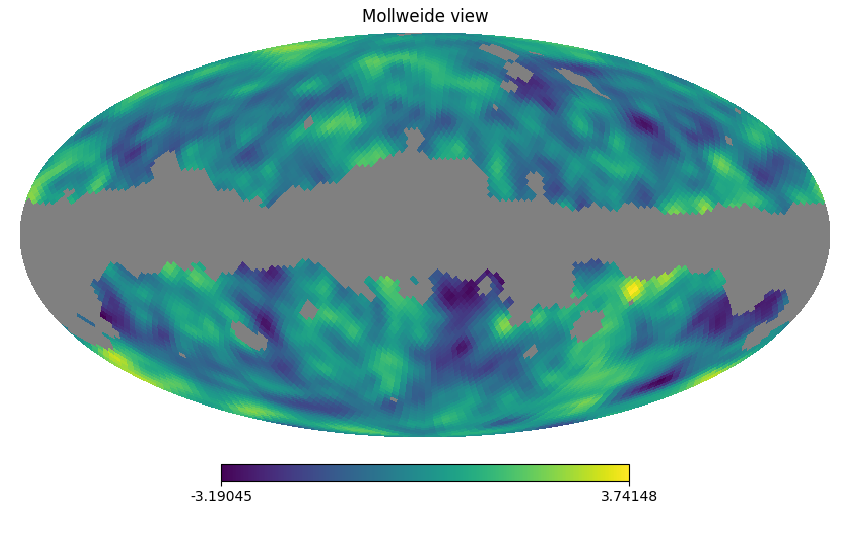} }
	\subfloat{\includegraphics[width=0.4\textwidth]{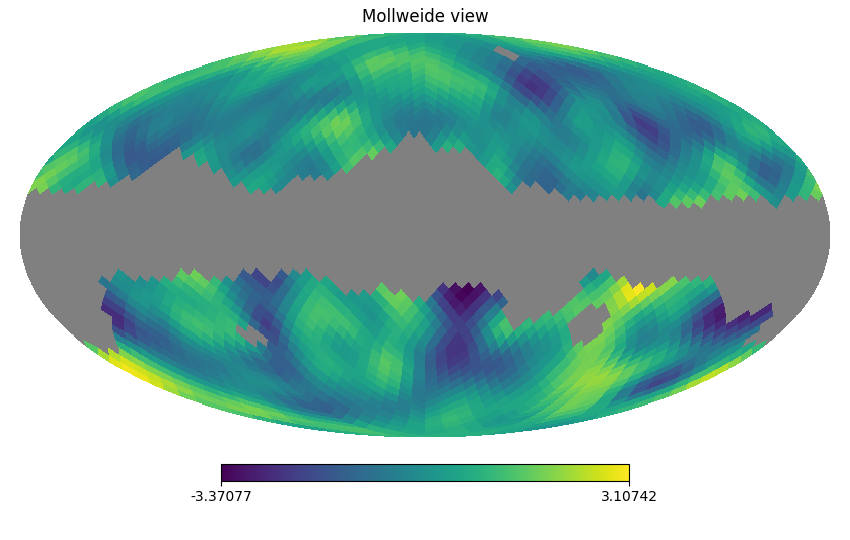} }\\
	\caption{A visualization of the degraded, smoothed and masked observational maps employed in this paper. The maps are degraded at $\Res = 2048,1024,512,256,128,64,32,16$, and smoothed at  $FWHM = 5', 10', 20', 40', 80', 160', 320', 640'$ respectively. The mask is degraded and smoothed at the same scales as the map, and subsequently thresholded at $0.9$ for re-binarization.}
	\label{fig:masked_maps}
\end{figure*}

\begin{table} 
	\begin{center} 
		\begin{tabular}{|r|r|r|} 
			\hline
			Resolution   &   FWHM(arcmin) &  \% unmasked  \\ \hline \hline
			2048			&	5 						&	77.9			\\ \hline 	
			1024			&	10 						&	76.9			\\ \hline 	
			512				&	20						&	75.6			\\ \hline 
			256				&	40						&	74.7			\\ \hline 
			128				&	80						&	73.6		\\ \hline 
			64				&	160						&	71.3		\\ \hline 
			32				&	320						&	68.8			\\ \hline 
			16				&	640						&	64.5			\\ \hline 
		\end{tabular} 
	\end{center} 
	\caption{Percentage of sky area covered by the unmasked regions for the various degraded resolutions and smoothing scales. The re-binarization threshold is for mask is set at $0.9$.} 
	\label{tab:unmasked_area} 
\end{table} 

The computational pipeline is composed of the steps described below in brief; see \cite{pranav2019b} for a substantially detailed description.

\medskip\noindent\textbf{Pre-processing.} We perform a range of pre-processing steps to the maps using the \texttt{HealPix} package \citep{healpix1}, which is also the format of the initial data. First, the CMB and the noise maps are added pixel-wise for each realization of the simulations. Subsequently, the observational and noise-added simulation maps are degraded and smoothed at a range of scales for a scale-dependent analysis.  The mask, presented in Figure~\ref{fig:mask}, which is a binary map, is also degraded and smoothed at the same resolution as the CMB maps, and subsequently thresholded at a value of $0.9$ to make it binary again. Table~\ref{tab:unmasked_area} presents the percentage of sky covered by the unmasked region for the degraded resolution and associated smoothing scales analyzed in this paper. Figure~\ref{fig:masked_maps} presents a visualization of the degraded and smoothed maps, after applying the mask. Next, we compute the mean and standard deviation from the unmaksed pixels, for each realization and resolution, and rescale the maps, pixel-wise, by the standard deviation after subtracting the mean. As a final step, we assign the value infinity to the masked pixels (numerically simulated by a very large number). 

\medskip\noindent\textbf{Triangulation.} As a first step, we project pixels of the maps  projected onto $\Sspace^2$, and we triangulate this set of points in $\Rspace^3$. Taking the convex hull of this triangulation produces a triangulation of the point-set on $\Sspace^2$. This triangulation consists of $V = 12 \Res^2$ vertices, $3V - 6$ edges, and $2V - 4$ triangles, where $\Res$ is the resolution parameter in \texttt{HealPix} format. It is the input to all downstream computations, and represents the temperature field, $f \colon \Sspace^2 \to \Rspace$, by storing the temperature value at each vertex \citep{pranav2019b}. We assume a piece-wise linear interpolation along higher dimensional simplices. 

\medskip\noindent\textbf{Upper-star filtration.} Given the triangulation $K$ constructed in the previous step, we order its simplices such that $\ssx$ precedes $\tsx$ if (i) $f(\ssx) > f(\tsx)$ or (ii) $f(\ssx) = f(\tsx)$ and $\dime{\ssx} < \dime{\tsx}$, in which $f(\ssx)$ is the minimum temperature value of the vertices of $\ssx$. Any ordering that satisfies (i) and (ii) is called an
\emph{upper-star filter} of $K$ and $f$. The corresponding \emph{upper-star filtration} consists of all
prefixes of the filter, each representing an excursion set of $f$. 

\medskip\noindent\textbf{Persistence computation.} We construct a boundary matrix from the upper-star filtration of $K$. Writing $\ssx_1, \ssx_2, \ldots, \ssx_n$ for the sorted simplices of the upper-star filtration, the  boundary matrix $\partial [1..n, 1..n]$ is defined by  $\partial [i,j] = 1$, if $\ssx_i$ is a face of $\ssx_j$ and $\dime{\ssx_i} = \dime{\ssx_j} - 1$, and $\partial [i, j] = 0$, otherwise. Computing the persistence birth death pair from this ordered boundary matrix involves reduction of the columns to the \emph{lowest-j} form. We resort to reduction from left to right, and a column of the matrix is reduced if if it is zero or its lowest $1$ has only $0$s in the same row to its left.  Each column with a unique \emph{lowest-j} contributes to the birth death pair of the persistence diagram, where the index of the birth and death simplices are precisely the row and column indices of the \emph{lowest-j}. Ranks of homology groups relative to the mask are inferred from the persistence diagrams by setting the vertices belonging to the mask at $+\infty$ \citep{pranav2019b}:

\begin{align}
\relBetti{0}  &= \# \{[b,d) \in Dgm_0(\Excursion \cup \Mask) \mid  +\infty > b \geq \nu > d \} ; \\ \nonumber
\relBetti{1}  &= \# \{[b,d) \in Dgm_0(\Excursion \cup \Mask) \mid  +\infty = b > d \geq \nu \} \\ \nonumber
&+ \# \{[b,d) \in Dgm_1(\Excursion \cup \Mask) \mid  +\infty > b \geq \nu > d \} ;\\ \nonumber
\relBetti{2} &= \# \{[b,d) \in Dgm_1(\Excursion \cup \Mask) \mid  +\infty = b > d \geq \nu \} \\ \nonumber
&+ \# \{[b,d) \in Dgm_2(\Excursion \cup \Mask) \mid  +\infty > b \geq \nu > d \} .
\end{align}

\medskip\noindent\textbf{Tests for determining statistical significance.} 
\label{sec:stat}
Our main aim is comparing the observational CMB maps with the null-hypothesis simulation maps, which assume isotropy, homogeneity and Gaussianity. Let $\x_i \in \R^m$, $i=1,\ldots,n$, be a sample of i.i.d.\ $m$-dimensional  vectors,  drawn from a distribution $G$. Let $\y \in \R^m$ be another sample point, assumed to be drawn from a distribution $F$. We wish to test the (null) hypothesis that $F=G$, and shall give the test results in terms of \emph{$p$-values}, that compute the probability that $\y$ is `consistent' with this hypothesis. We employ two different statistical tests, which we briefly describe below. 

\medskip\noindent\emph{Mahalanobis distance.} The first is the parametric \emph{Mahalanobis distance}, or the familiar $\chi^2$ test \citep{mahalanobis}. If $G$ is assumed to be Gaussian and $n$ is large,  then under the hypothesis that $G=F$ the squared Mahalanobis distance is approximately distributed as a   $\chi^2$ distribution with $m$ degrees of freedom. Thus the corresponding  $p$-value is

\begin{equation}\label{eq:Mahal-p}
p_{\rm Mahal}(\y) = P[\chi^2_m > d^2_{\rm Mahal}(\y)].
\end{equation}

\medskip\noindent\emph{Tukey depth.} The second method is a non-parametric test based on the \emph{Tukey depth} \citep{depth}. It a general metric  for identifying outliers  in a flexible manner and in a non-parametric setting, making no assumptions on the structure of $F$ and $G$. Let $\mathbf{z}$ be any point in $\mathbb R^m$.  then the half-space depth $d_{\rm dep}(\mathbf{z}; \x_1,\ldots,\x_n)$ of $\mathbf{z}$ within the sample of the $\x_i$ is the smallest fraction of the $n$ points $\x_1,\ldots,\x_n$ to either side of any hyperplane passing through $\mathbf{z}$.  Points that have the same depth constitute a non-parametric estimate of the  isolevel contour of the distribution $F$. We first compute $d_j = d_{\rm dep}(\x_j; \x_1,\ldots,\x_n)$ for every point $\x_j$, $j=1,\ldots,n$, yielding an empirical distribution of depth. The $p$-value of  $\y$ is computed as the proportion of points whose depth is lower than that of $\y$: 

\begin{equation}\label{eq:hsd-p}
p_{\rm dep} (\y)  =  \#\{j \mid d_j > d_{dep}(y) \} / n
\end{equation}

\section{\texttt{FFP10} results}

In this section, we present the graphs and the table of $p$-values corresponding to the \emph{Planck 2018} Data release 3 (DR3)
 dataset. We base the significance of our results on the accompanying $300$ \texttt{FFP10} simulations.
\begin{figure}
	\centering
	\subfloat{\includegraphics[width=0.7\textwidth]{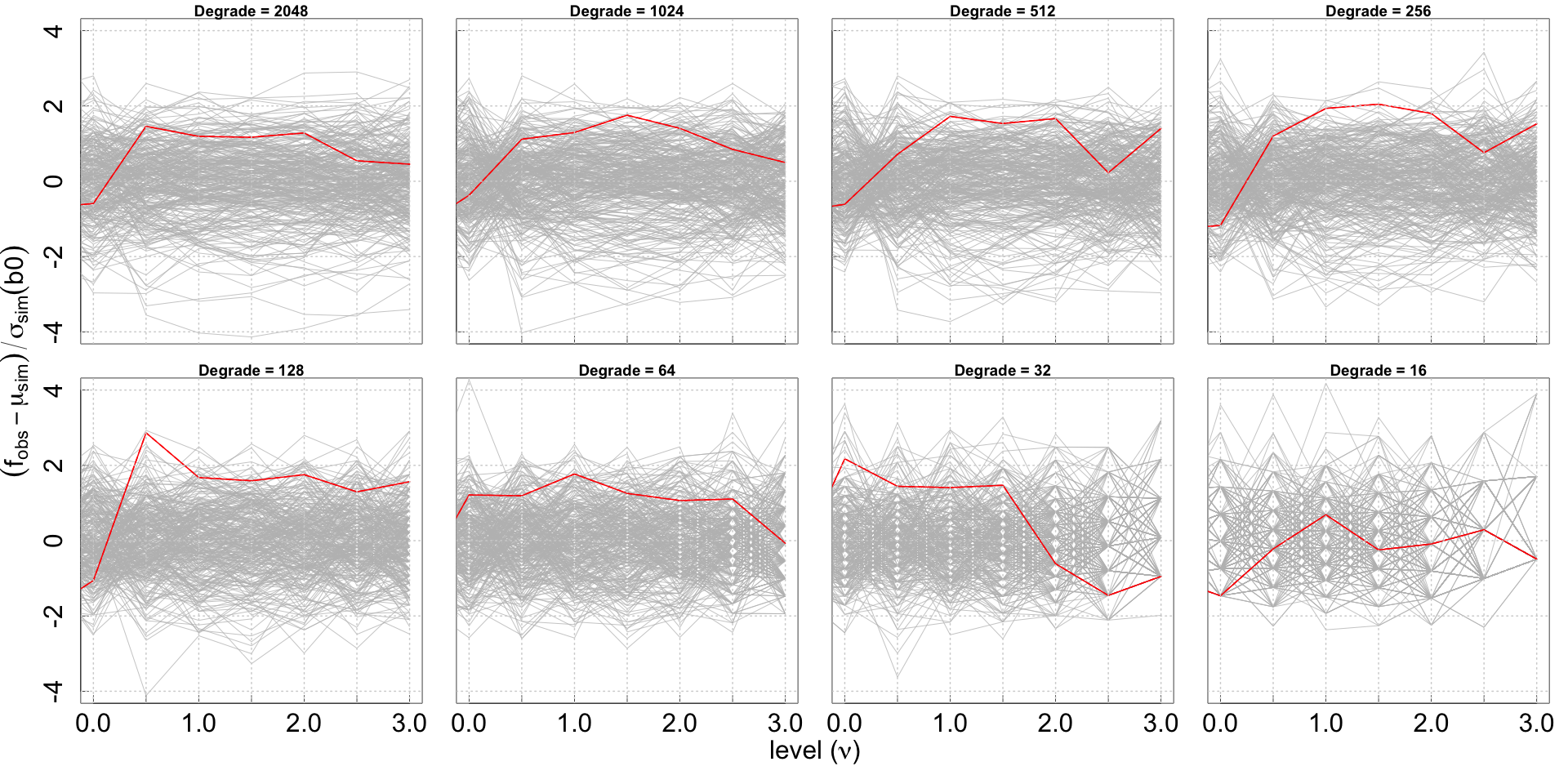} }\\
	\subfloat{\includegraphics[width=0.7\textwidth]{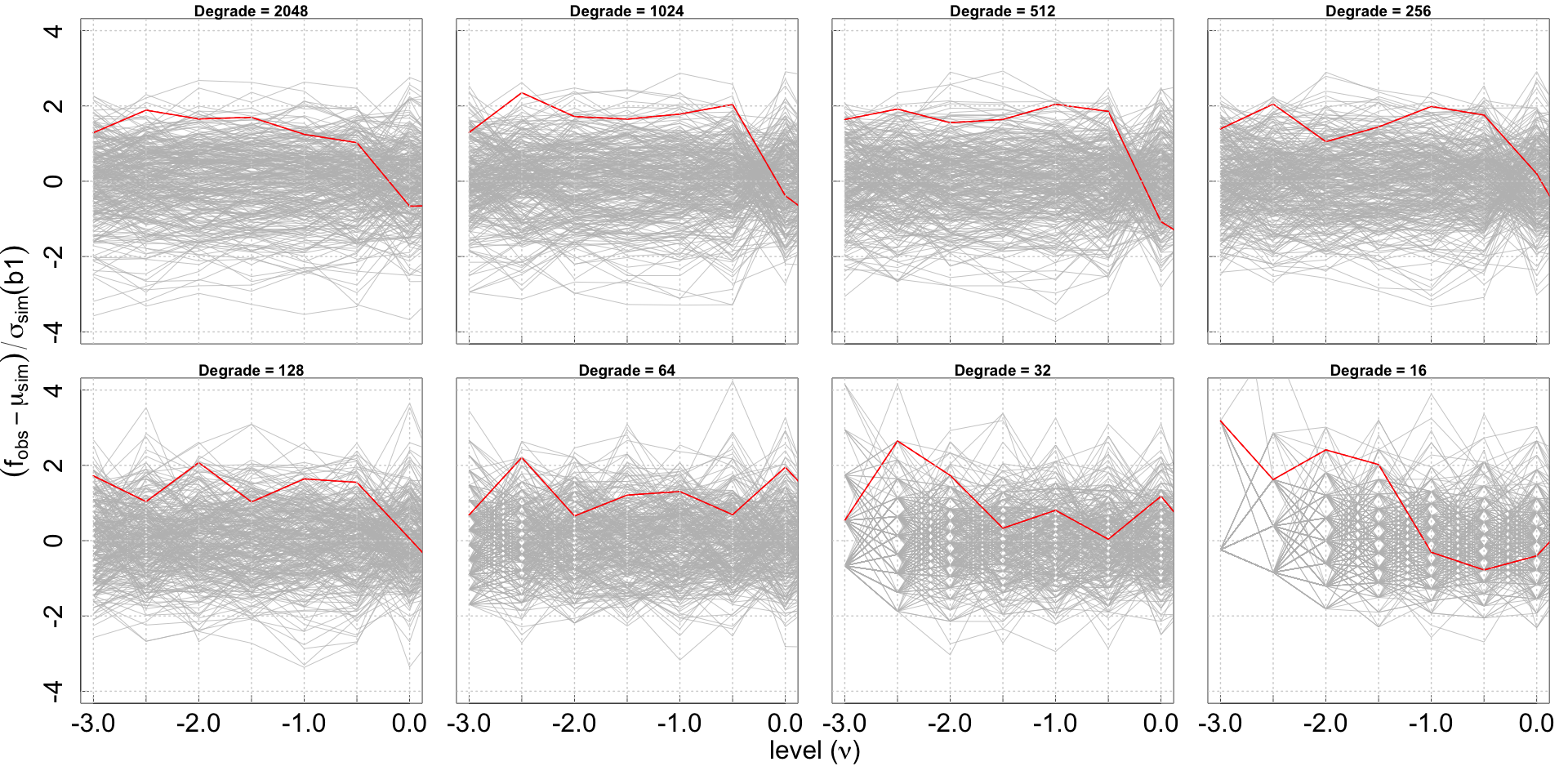} }\\
	\subfloat{\includegraphics[width=0.7\textwidth]{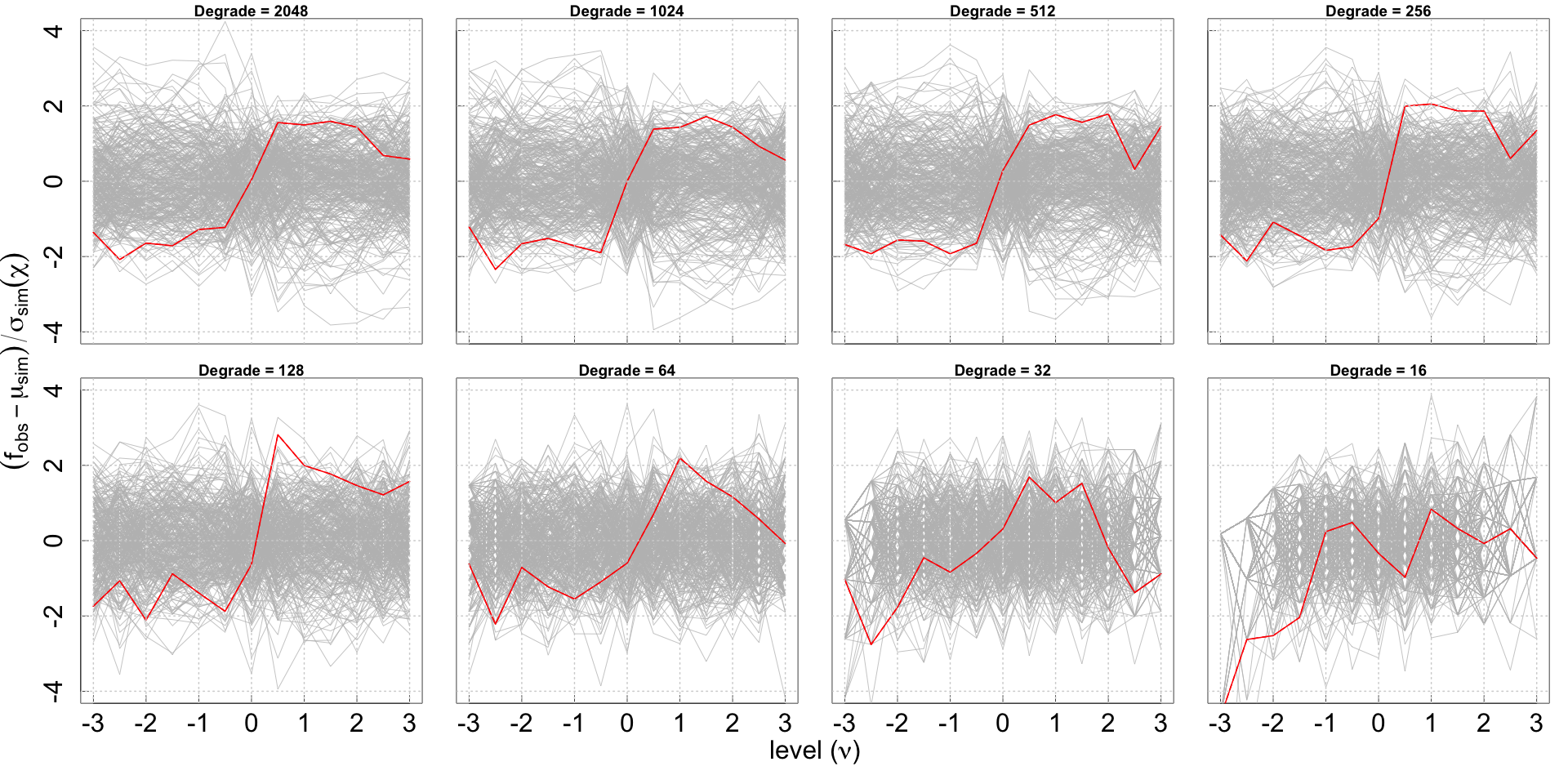} }\\
	\caption{Significance of difference for $\relBetti{0}$, $\relBetti{1}$ and $\relEuler$ for the \texttt{FFP10} dataset.}
	\label{fig:nrm_graph_ffp}
\end{figure}

\begin{table}
	\tabcolsep=0.09cm
	\centering
	\subfloat{\reltabffp}\\
	\caption{Table displaying the two-tailed $p$-values for relative homology obtained from parametric (Mahalanobis distance) and non-parametric (Tukey depth) tests, for different resolutions and smoothing scales for the \texttt{FFP10} dataset. The last entry is the $p$-value for the summary statistic computed across all resolutions. Marked in boldface are $p$-values $0.05$ or smaller.} 
	\label{tab:ffp10_degrade-pvalues}
\end{table}

\end{appendix}

\end{document}